\documentclass[prb,aps,amssymb,twocolumn,showpacs,floatfix]{revtex4}
\usepackage{graphics}
\usepackage{epsfig} 
\usepackage{dcolumn}
\usepackage{amsmath}
\begin{document}
\newcommand\dd{{\operatorname{d}}}
\newcommand\sgn{{\operatorname{sgn}}}
\def\Eq#1{{Eq.~(\ref{#1})}}
\def\Ref#1{(\ref{#1})}
\newcommand\e{{\mathrm e}}
\newcommand\cum[1]{  {\Bigl< \!\! \Bigl< {#1} \Bigr>\!\!\Bigr>}}
\newcommand\vf{v_{_\text{F}}}
\newcommand\pf{p_{_\text{F}}}
\newcommand\ef{{\varepsilon} _{\text{\sc f}}}
\newcommand\zf{z_{_\text{F}}}
\newcommand\zfi[1]{{z_{_\text{F}}}_{#1}}
\newcommand\av[1]{\left<{#1}\right>}
\def\det{{\mathrm{det}}}
\def\Tr{{\mathrm{Tr}}}
\def\Li{{\mathrm{Li}}}
\def\tr{{\mathrm{tr}}}
\def\im{{\mathrm{Im}}}
\def\Texp{{\mathrm{Texp}\!\!\!\int}}
\def\antiTexp{{\mathrm{\tilde{T}exp}}\!\!\!\int}
\title{Bosonization of one dimensional fermions out of equilibrium}

\author{D. B. Gutman$^{1,2,3}$, Yuval Gefen$^4$, and A. D. Mirlin$^{5,2,3,6}$}
\affiliation{
\mbox{$^1$The department of Physics, Bar Ilan University, Ramat Gan 52900,
Israel }\\
\mbox{$^2$Institut f\"ur Theorie der kondensierten Materie,
Karlsruhe Institute of Technology, 76128 Karlsruhe, Germany}\\
\mbox{$^3$DFG Center for Functional Nanostructures,
Karlsruhe Institute of Technology, 76128 Karlsruhe, Germany}\\
\mbox{$^4$Dept. of Condensed Matter Physics, Weizmann Institute of
  Science, Rehovot 76100, Israel}\\
\mbox{$^5$Institut f\"ur Nanotechnologie, Karlsruhe Institute of Technology,
 76021 Karlsruhe, Germany}\\
\mbox{$^6$Petersburg Nuclear Physics Institute, 188300 St.~Petersburg, Russia}
}

\date{\today}

\begin{abstract}
Bosonization technique for one-dimensional fermions out
of equilibrium is developed in the framework of the Keldysh action formalism.
We first demonstrate how this approach is
implemented for free fermions and for the problem of non-equilibrium
Fermi edge singularity. We then employ the technique to study an interacting
quantum wire attached to two electrodes with arbitrary energy distributions.
The non-equilibrium electron Green functions, which can be measured
via tunneling spectroscopy technique and carry the information
about energy distribution, zero-bias anomaly, and dephasing,
are expressed in terms of
functional determinants of single-particle ``counting'' operators.
The corresponding time-dependent scattering phase is found to be intrinsically
related to ``fractionalization'' of electron-hole
excitations in the tunneling process and at boundaries with leads.
Results are generalized to the case of spinful particles as well to
Green functions at different spatial
points (relevant to the problem of dephasing in Luttinger liquid
interferometers). For
double-step distributions, the dephasing rates are
oscillatory functions of the interaction strength.
\end{abstract}

\maketitle
\section{Introduction}
One-dimensional (1D) interacting fermionic systems show remarkable
physical properties.  The electron-electron interaction
manifests itself in a particularly dramatic way in 1D systems, inducing a
strongly correlated electronic state -- Luttinger liquid (LL)
\cite{Gogolin,stone,giamarchi,maslov-lectures,Delft}.
A paradigmatic experimental realization of quantum
wires are carbon nanotubes \cite{zba-carbon-nanotubes}; for a recent
review see Ref.~\onlinecite{nanotubes}.
Further realizations encompass semiconductor\cite{semiconductor}, metallic
\cite{metallic} and polymer nanowires\cite{polymer}, as well
as quantum Hall edges\cite{Hall,MachZehnder}.

While equilibrium LL have been extensively explored, there is
currently a growing interest in non-equilibrium phenomena on nanoscale
and, in particular, in non-equilibrium properties of quantum wires.
In a recent experiment \cite{Chen09}  the tunneling spectroscopy
of a biased LL conductor has been performed (see also a related work on
carbon nanotube quantum dots \cite{dirks09}). A similar approach was
used to study experimentally non-equilibrium quantum Hall edges
\cite{altimiras09}. Quite generally, the tunneling spectroscopy
technique allows one to measure the non-equilibrium Green
functions $G^\gtrless(\tau)$. Analogous experiments \cite{Pothier}
have been carried out earlier in order to study energy distribution
function and inelastic relaxation processes in quasi-one-dimensional
diffusive metallic samples. The interpretation of the results for a
metallic sample is based on the Fermi liquid theory, and, in
particular, on a kinetic equation for a quasi-particle distribution
function. In fact, even in that case, careful analysis requires
taking into account non-equilibrium dephasing processes \cite{GGM}
which lead to additional broadening of the measured Fermi-edge
structures in the tunneling current. In the case of strongly
correlated,  non-Fermi-liquid systems (such as LL) out of equilibrium,
the situation is much more complex. In this situation not only a
quantitative theoretical analysis of $G^\gtrless$, but even the very notions
of quasiparticle energy distribution and dephasing, become highly
non-trivial. The goal of the presented  work is to construct  a
corresponding theory. To
achieve this goal, we develop a formalism of non-equilibrium (Keldysh)
bosonization. While we consider systems of 1D interacting electrons in
this work, we expect that it will be an important step in
understanding the properties of a broader class of systems---non-equilibrium
quantum fluids in low-dimensions. This includes, in particular,
systems of cold atoms, with either fermionic or bosonic statistics.

The structure of the present paper is as follows:

In  Sec.~\ref{sec_setups} we discuss possible experimental realizations
of a non-equilibrium LL.

In  Sec.~\ref{sec_free_fermions}
we develop a bosonization technique for  non-interacting electrons
away from equilibrium. Working within the Keldysh non-equilibrium
formalism, we derive the action of the
bosonized theory. While this action is quadratic at equilibrium (which
is the essence of conventional bosonization), it now includes
arbitrary powers in the bosonic fields. We demonstrate how this action
can be used to express the Green function of non-interacting
fermions in terms of a Fredholm functional determinant of a single-particle
``counting'' operator (which is of Toeplitz type).
We further discuss the relation between this
problem and that of counting statistics.
Specifically, our result is expressed in terms
of the determinant at the value of the phase (``counting field'')
$\lambda=2\pi$. On the other hand, the counting statistics at this
point is trivial, in view of charge quantization. We show that the
difference between the determinants used for expressing the Green
functions and those used for counting statistics results from
different continuations (analytic vs. periodic) of the
functional determinant beyond the non-analyticity point $\lambda=\pi$.

In Sec.~\ref{sec_FES} we apply our technique to the problem of
Fermi edge singularity (FES) out of equilibrium. We show that
non-equilibrium FES Green function is expressed in terms of the same
functional determinant but with a shifted value of the argument,
$\lambda = 2(\pi - \delta_0)$ , where $\delta_0$ is the scattering
phase on the core hole. Comparing our results for this problem with those
obtained earlier\cite{Abanin}, we establish useful
identities between  Fredholm determinants
of counting operator at values of the counting field $\lambda$ differing by
$2\pi$.

In  Sec.\ref{sec_interaction} our  formalism  is
extended to interacting fermions in a quantum wire.
First, we analyze the problem of tunneling spectroscopy
of a non-equilibrium LL in the case of spinless fermions. We
demonstrate that the non-equilibrium Green functions are expressed in
terms of products of single-particle Fredholm determinants. The
corresponding values of the counting fields are shown to be related to
``fractionalization'' of particle-hole excitations created during
the tunneling process, as well as at the boundaries with non-interacting
leads. Our results for $G^\gtrless$ contain all information about
single-particle properties of the system, including tunneling density
of states, energy distribution, and dephasing.
We find, in particular, that the dephasing rate
oscillates as a function of the interaction strength (LL parameter
$K$), vanishing at certain values of $K$.
At the end of the section we generalize the consideration to the case
of spinful fermions, as well to Green functions at different spatial
points (which is relevant to the problem of dephasing in LL
interferometers).

Section \ref{Summary} includes a summary of our results as well as  prospects
for future work.

Some of results of this work were presented in a Letter,
Ref.~\onlinecite{GGM_2009_short}.

\section{Nonequilibrium Luttinger liquid: Setups}
\label{sec_setups}

In this section we specify the class of problems to be considered and
discuss possible experimental setups. We assume that electrons with
distributions functions $n_\eta(\epsilon)$ ($\eta=R,L$ labels right-
and left-movers) are injected into a LL wire from two non-interacting
electrodes. It is convenient to model the electrodes as non-interacting 1D
systems, so that the whole structure is a wire with spatially
dependent interaction that switches on near the points $x=\pm L/2$,
see Sec.~\ref{sec_interaction} for detail.

It is worth noting that we assume the absence of electron backscattering due to
impurities inside the LL wire. When present in sufficient amount (so
that one can speak about a disordered LL), such impurities strongly affect
the electronic properties of a LL wire. Specifically,
they induce diffusive dynamics at
sufficiently high temperature $T$ and  localization phenomena
proliferating with lowering $T$
(Ref.~\onlinecite{Apel82,Giamarchi,GMP}), as well as
inelastic processes \cite{Bagrets1}. We also neglect the nonlinearity
of the electron dispersion whose influence on spectral and kinetic
properties of 1D electrons was recently studied in
Refs.~\onlinecite{khodas}, \onlinecite{Matveev}.

We discuss now possible experimental realizations of the problem. The
simplest way to take the system out of equilibrium is to apply a
voltage between two electrodes, so that the incoming
distribution functions have different chemical potentials,
$\mu_L-\mu_R=eV$, but equal temperatures, $T_R=T_L=T$, see
e.g. Ref.~\onlinecite{Chudnovsky}. However, in the case
of a LL this situation is almost identical to the equilibrium
one, in view of the absence of electron backscattering.
Indeed, the bosons remain at equilibrium, so that the usual
bosonization technique (within Matsubara formalism) can be applied. The only
non-equilibrium effect will be a simple shift of the chemical
potential of left- movers as compared to that of the right-movers.

\begin{figure}
\includegraphics[width=0.8\columnwidth,angle=0]{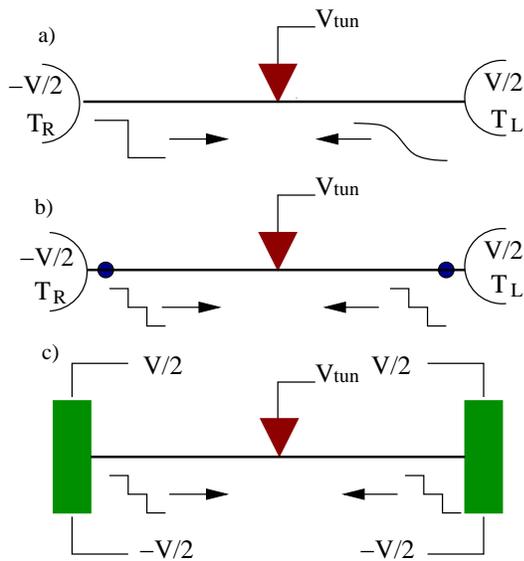}
\caption{Schematic view of experimental setups for tunneling
spectroscopy of a LL out of equilibrium: (a) ``partially
non-equilibrium'' setup, with distribution functions
$n_\eta(\epsilon)$ of Fermi-Dirac form but with different
temperatures; (b), (c) ``fully non-equilibrium'' setups characterized by
double-step distribution functions $n_\eta(\epsilon)$ of electrons
injected into the LL wire.}
\label{setups}
\end{figure}

A generalization of this setup that does yield
a non-trivially non-equilibrium LL is shown in Fig.~\ref{setups}a.
A long clean LL is adiabatically coupled to two electrodes
with different potentials, $\mu_L - \mu_R = eV$
and different temperatures $T_L$, $T_R$. (A particularly interesting
situation arises when one of temperatures is much
larger than the other, e.g., $T_L=0$ and $T_R$ finite, so that
non-equilibrium effects are most pronounced.)
This model has been investigated in our previous works,
Refs.~\onlinecite{GGM2008,GGM_2009}. While showing genuinely non-equilibrium
effects (in particular, energy redistribution of electrons), this
model, when treated in the framework of Keldysh bosonization
formalism, is characterized by a Gaussian action.
For this reason, we termed this setup
``partially non-equilibrium'' in Ref.~\onlinecite{GGM2008}. We will verify in
Sec.\ref{sec_interaction} that the results of the present work
(pertaining to full non-equilibrium) reduce to
those obtained earlier (partial non-equilibrium) in the case when both $n_R$
and $n_L$ are taken to be Fermi-Dirac functions.

The focus of this work is generic non-equilibrium situations, when
at least one of the functions $n_\eta$ is not of the Fermi-Dirac form.
Such situations naturally arise when electrons injected into a LL
wire represent juxtaposition of particles originating from
reservoirs with different chemical potentials and mixed by impurity
scattering. Two possible realizations of such devices are shown in
Fig.~\ref{setups}b,c.
In the first case, Fig.~\ref{setups}b, the mixture of left and right
movers coming from reservoirs with $\mu_L \ne \mu_R$
is caused by
impurities which are located in the  non-interacting part of the wires
\cite{Ngo}.
In the second setup, Fig.~\ref{setups}c, the LL wire is
attached to two thick metallic wires which are themselves
biased. We assume that those electrodes are diffusive but
sufficiently short, so that energy equilibration there can be
neglected. As a result, a double-step energy
distribution is formed in the electrodes \cite{Pothier} and
``injected'' into the LL conductor.
Such double-step distributions are of particular interest for our
problem, as they are of the ``maximally non-equilibrium'' form.
The existence of multiple Fermi edges in the  distribution
functions ``injected'' from the electrodes renders the electron-electron
scattering processes \cite{GGM,Altland} which govern the non-equilibrium
dephasing rate $\tau_\phi$ (and thus the broadening of tunneling
spectroscopy characteristics) particularly important.

The question of non-equilibrium dephasing induced by
electron-electron scattering is
particularly intriguing in the case of a 1D system.
First,  energy relaxation is absent in a homogeneous LL system. Second,
recent analysis of dephasing in the context of weak localization and
Aharonov-Bohm oscillations has given qualitatively different results:
while the weak-localization dephasing rate vanishes in the limit of
vanishing disorder \cite{GMP},
the Aharonov-Bohm dephasing rate is finite in a clean LL
\cite{GMP,lehur}. In the case of a partially non-equilibrium setup
the tunneling spectroscopy
dephasing rate has a form similar to the equilibrium Aharonov-Bohm
dephasing rate \cite{GGM2008,GGM_2009}. As we show here, in the case
of double-step distributions dephasing acquires 
qualitatively distinct features; in particular, the dephasing rate
becomes an oscillatory function of the interaction strength.

Having described the problems to be addressed, we turn to the
corresponding formalism. It is instructive to develop it first for the
case of non-interacting fermions and then  ``turn on'' the
interaction.

\section{Free  fermions}
\label{sec_free_fermions}

In this section we develop a bosonization formalism for
the case of free fermions out of equilibrium. Specifically, we
consider non-interacting fermions with a given distribution function
$n(\epsilon)$ and derive the corresponding bosonic action. Using the
latter, we calculate the fermionic Green function. Clearly, the Green
function of non-interacting fermions is trivially obtained within the
fermionic formalism. However, the results of this section are not just
a complicated way to calculate a simple quantity. Rather, they will
play a crucial role for developing the bosonic formalism for
interacting systems studied in the remainder of the paper.

\subsection{Keldysh  action: From fermions to bosons}
\label{sec_keldysh_action}

Bosonization has been proved to be a very efficient tool for
tackling one dimensional problems at
equilibrium\cite{Gogolin,stone,giamarchi,maslov-lectures,Delft}, as it
maps a system of interacting fermions (LL) onto that of
non-interacting bosons. One can thus hope for similar advantages of
this approach  for  non-equilibrium problems as well.
The question though is whether  the
bosonization procedure can be generalized to  systems out of
equilibrium? As we show below, the answer is affirmative, yet substantial
modifications are required.

Quite generally, operator bosonization procedure consists of the
following steps:
(i) mapping between the Hilbert space of fermions and bosons;
(ii) construction of the bosonic Hamiltonian $H_B$ representing the
original fermionic Hamiltonian $H_F$  in terms of bosonic
 (particle-hole) excitations, i.e. density fields;
(iii) expressing fermionic operators in the bosonic language;
(iv) calculation of observables (Green functions) within the bosonized
formalism by averaging with
respect to the many body bosonic density matrix (${\rho}_B$).
Neither the Hilbert space nor the operators (including the
Hamiltonian) contain an information
regarding a state of the many-body system. Therefore, the first three steps
remain unchanged for a non-equilibrium situation.
The major modifications occur in the step (iv).
Indeed, at equilibrium  the fermionic density matrix
is expressed through the corresponding  Hamiltonian as
$\rho_F=\exp(-H_F/T)$, implying that the same relation
holds in the bosonized theory, $\rho_B=\exp(-H_B/T)$,
which makes averaging with respect to $\rho_B$ straightforward.
Out of equilibrium this is not so anymore: a one-particle density
matrix corresponding to a non-equilibrium occupation $n(\epsilon)$ of
fermionic states translates into a complicated density matrix of
bosons, which does not allow the application  of  Wick theorem. This poses a
major difficulty in  bosonizing fermionic problems away from equilibrium and, as
we see below, results in a non-gaussian action of the bosonized theory.

To construct the effective bosonic theory, we start with
the fermionic description.
Within the LL model, the electron field is decoupled into a sum of left-
and right-moving terms,
\begin{equation}
\label{fermi_operator1}
\psi(x,t)=\psi_R(x,t)e^{ip_Fx}+\psi_L(x,t)e^{-ip_Fx}\,,
\end{equation}
where $p_F$ is the Fermi momentum. The Hamiltonian of the system reads
\begin{equation}
\label{Hamiltonian_free}
H_0=-iv\int dx \left(\psi_R^\dagger\partial_x\psi_R-
\psi^\dagger_L\partial_x\psi_L\right)\,,
\end{equation}
where $v$ is the electron velocity.
The bosonic representation  for fermionic operators has the form
\cite{Gogolin,stone,giamarchi,maslov-lectures,Delft,Klein_factors}
\begin{align}
\label{fermi_operator2}
\psi_{\eta}(x)\simeq\left(\frac{\Lambda}{2\pi v}\right)^{1/2}
e^{\eta ip_Fx}
e^{i\phi_\eta(x)}\,,
\end{align}
where $\Lambda$ is an ultraviolet cut-off. The bosonic fields
$\phi_\eta(x)$ are related to the density of electrons (given by
$\rho_{\eta}(x) = \psi_\eta^\dagger(x)\psi_\eta(x)$ in the fermionic language)
as
\begin{align}
\rho_{\eta}(x)=\frac{\eta}{2\pi}\partial_x \phi_{\eta}\,,
\end{align}
and obey the commutation relations
\begin{equation}
[\phi_R(x),\phi_R(x')]=-[\phi_L(x),\phi_L(x')]=i\pi\sgn(x-x')\,.
\end{equation}
We use the convention that in formulas $\eta$ should be understood  as
$\eta =\pm1$ for right/left moving electrons.
The bosonized Hamiltonian is expressed in terms of density fields in
the following way:
\begin{equation}
\label{Hamiltonian_free_bosons}
H_0=\pi v\int dx \left(\rho_R^2+\rho_L^2\right)\,.
\end{equation}

We turn now to the Lagrangian formalism. Since we deal with
a non-equilibrium situation, the system is characterized by an action defined
on the Keldysh contour\cite{Kamenev},
\begin{align}
\label{free_fermions}
S_0[\psi]=\int_cdt\int dx \sum_{\eta=R,L}
\psi^\dagger_\eta i\partial_\eta \psi_\eta
\end{align}
where $\psi,\psi^\dagger(t,x)$ are fermionic fields, and
$\partial_{R,L} = \partial_t\pm v\partial_x$.
To generate correlation functions, it is convenient to introduce a
source term.
\begin{align}
S_V[\psi]=\int_cdt \int dx  V_\eta(x,t)\psi_\eta^\dagger(x,t)\psi_\eta(x,t)\,.
\end{align}
The field components on the upper branch and lower are denoted by $+$
and $-$ respectively.
It is convenient to perform a rotation
in Keldysh space\cite{Kamenev}, thus
decomposing fields into classical and quantum components (the latter
being denoted by a bar),
\begin{eqnarray}&&
V_\eta, \bar{V}_\eta=(V_{+,\eta} \pm V_{-,\eta})/\sqrt{2}\,, \\&&
\rho_\eta,\bar{\rho}_\eta
=(\rho_{+,\eta} \pm \rho_{-,\eta})/\sqrt{2}\,,
\\ &&
\psi_\eta,\bar{\psi}_\eta = (\psi_{+,\eta} \pm \psi_{-,\eta})/\sqrt{2}\,,
\\ &&
\psi^\dagger_\eta,\bar{\psi}_\eta^\dagger = (\psi^\dagger_{+,\eta} \mp
\psi^\dagger_{-,\eta})/\sqrt{2}\,.
\end{eqnarray}
In these notations,
the density correlation functions are encoded in the generating function
\begin{equation}
\label{gen_funct}
{ Z}_\eta[V_\eta,\bar{V}_\eta]=\langle
\exp\{iV_\eta\bar{\rho}_\eta+i\bar{V}_\eta\rho_\eta\}\rangle_{S_0}\,.
\end{equation}
The calculation of the partition function can be performed
in either the fermionic or the bosonic description.
In the fermionic language it can be readily done by
evaluating a Gaussian integral over the Grassman variables,
\begin{eqnarray}
\label{gen_funct2}&&
{ Z}_{\eta}[V,\bar{V}]=
\det[1+ G_{\eta 0} (\sigma_0V+
\sigma_1\bar{V})/\sqrt{2}]\,,
\end{eqnarray}
where $\sigma_0$ and $\sigma_1$ are the unit matrix and the first Pauli
matrix in the Keldysh space, and $G_{\eta 0}$ is the Keldysh Green function of
free chiral fermions, which has the following matrix structure:
\begin{eqnarray}
\label{green-function-phi}
&&G_{\eta 0}=
\begin{pmatrix}
     G^r_{\eta 0} & G^K_{\eta 0}\\
      0 & G^a_{\eta 0}
    \end{pmatrix}\:\!.
\label{green-function-free}
\end{eqnarray}
Here $G^a_{\eta,0}$, $G^r_{\eta,0}$ and $G^K_{\eta,0}$ are advanced, retarded and
Keldysh components,
\begin{eqnarray}
\label{definitions}
&& G_{\eta 0}^{r,a}(\epsilon,p) = 1 / (\epsilon-\eta vp\pm i0)\,; \\
&& G_{\eta 0}^K(\epsilon,p) = [1-2n_\eta(\epsilon)] [G_{\eta 0}^r(\epsilon,p)-
G_{\eta 0}^a(\epsilon,p)]\,.
\end{eqnarray}

We expand now the generating functional
(\ref{gen_funct2}) in powers of the source fields $V_\eta$, $\bar{V}_\eta$.
For higher-dimensional systems this would generate all terms of the
type $V_\eta^n\bar{V}_\eta^m$. In 1D the situation is different. Specifically,
in an equilibrium 1D system only terms up to  second order
($V_\eta\bar{V}_\eta$ and $\bar{V}_\eta^2$) are generated, which forms
the basis of conventional bosonization. Out of equilibrium, this is not true
anymore: terms of higher orders are generated as well, and the theory
becomes non-gaussian. What is crucial, however, is that all
higher-order terms are of the type $\bar{V_\eta}^n$, i.e. they do not
depend on $V_\eta$. We will prove this statement in Sec.~\ref{sec_DL} and
\ref{sec_functional_det} below.

The generating functional has thus the structure
\begin{align}
\label{identity}
{ Z}_{\eta}[V,\bar{V}]=
\exp\left(
-iV_{\eta}\Pi^{a}_\eta \bar{V}_{\eta}
+
\sum_{n=2}^\infty\frac{i^n}{n!}\bar{V}_{\eta}^n {\cal S}_{n,\eta} \right)\,,
\end{align}
where  ${\cal S}_{n,\eta}$ is the $n$-th order irreducible vertex function,
\begin{eqnarray}
\label{vertex_functions}
&& {\cal S}_{n,\eta}(x_1,t_1;\ldots; x_n,t_n)\nonumber \\
 && \qquad =-i^n  \sum_{\rm perm.} \Tr_K
G_{\eta  0}(x_1,t_1;x_{i_2},t_{i_2}){\sigma_1\over\sqrt{2}} \nonumber \\
&& \qquad
\times G_{\eta  0}(x_{i_2},t_{i_2};x_{i_3},t_{i_3}){\sigma_1\over\sqrt{2}}
\times \ldots \nonumber\\  && \qquad \times
G_{\eta  0}(x_{i_n},t_{i_n};x_1,t_1){\sigma_1\over\sqrt{2}}
\,.
\end{eqnarray}
The multiplication in Eq.~(\ref{identity}) and analogous formulas
below should be understood in the matrix sense with respect to the
coordinates,
\begin{eqnarray}
V_{\eta}\Pi^{a}_\eta
\bar{V}_{\eta}
&=& \int [dx][dt]
V_{\eta}(x_1,t_1)\\
&\times & \Pi^{a}_\eta(x_1,t_1;x_2,t_2)\bar{V}_{\eta}(x_2,t_2)\,,
\nonumber \\
\bar{V}_\eta^n{\cal S}_{n,\eta} &=&
\int [dx][dt]
\bar{V}_{\eta}(x_1,t_1)\dots \bar{V}_{\eta}(x_n,t_n)\nonumber \\
& \times & {\cal S}_{n,\eta}(x_1,t_1;\dots;x_n,t_n)\,,
\end{eqnarray}
where $\int[dx][dt]$ implies integration over all spatial and time coordinates.
The summation in Eq.~(\ref{vertex_functions}) goes over $(n-1)!$
permutations $\{i_2,i_3,\ldots,i_n\}$ of the set of indices
$\{2,\ldots,n\}$ (labeling the space-time coordinates),  and ${\rm
  Tr}_K$
denotes the
trace over Keldysh indices. Clearly, after integration with
$\bar{V}_{\eta}$ fields in Eq.~(\ref{identity}) all the $(n-1)!$ terms
of the sum in Eq.~(\ref{vertex_functions}) yield equal contributions,
so that the total combinatorial factor is  $(n-1)!/n! = 1/n$,
as should be in the expansion of the logarithm. We have
chosen to define the vertex function in the symmetrized form
(\ref{vertex_functions}) [and to introduce the corresponding factor
$1/(n-1)!$  in Eq.~(\ref{identity})], since  ${\cal
  S}_{n,\eta}(x_1,t_1;\ldots; x_n,t_n)$ are
then equal to irreducible density correlation functions
$\langle\langle \rho(x_1,t_1)\ldots \rho(x_n,t_n)\rangle\rangle$.

The quadratic part of the generating functional (\ref{identity}) is
determined by the polarization operator of fermions,
\begin{eqnarray}
\Pi_\eta=\begin{pmatrix}
     0 & \Pi^a_{\eta}\\
      \Pi^r_\eta & \Pi^K_\eta
    \end{pmatrix},
\label{Polarization_operator}
\end{eqnarray}
with the retarded, advanced, and Keldysh components given by
\begin{eqnarray}
&& \Pi^{r,a}_\eta(\omega,q)= \frac{1}{2\pi}\frac{\eta q}{\eta v q -
  \omega \mp i0},
\\&&
\Pi^K_\eta(\omega,q)
=[\Pi^r_\eta(\omega,q)-\Pi^a_\eta(\omega,q)]B_\eta(\omega)\,.
\label{Pi}
\end{eqnarray}
Here the function
\begin{equation}
\label{e7}
B_\eta(\omega)=\frac{1}{\omega}\int_{-\infty}^{\infty} d\epsilon\:
n_\eta(\epsilon)\:[2-n_\eta(\epsilon-\omega)-n_\eta(\epsilon+\omega)],
\end{equation}
governs the distribution function $N_\eta(\omega)$ of electron-hole
excitations  moving with velocity $v$ in direction $\eta$,
$B_\eta(\omega) = 1 + 2 N_\eta(\omega)$.
At equilibrium,
\begin{equation}
B_\eta(\omega) = B_{\rm eq}(\omega) = 1+ 2N_{\rm eq}(\omega)=\coth\left(
\frac{\omega}{2T}\right)\,,
\end{equation}
where $N_{\rm eq}(\omega)$ is the Bose distribution.
By construction, the  second order density-correlation function
 ${\cal S}_{\eta,n=2}$ in Eq.~(\ref{identity}) is equal to the Keldysh
 component $\Pi_{\eta}^K$ of the polarization operator (times $-i$).

In order to bosonize the theory, we should find a bosonic counterpart
of the action $S_{0\eta}$ that reproduces the generating functional
(\ref{identity}). According to Eq.~(\ref{gen_funct}), we have
\begin{align}
\label{bosonic_action}
\exp\left(iS_{0\eta}[\bar{\rho}_\eta,\rho_\eta]\right)=
\int {\cal D}V_\eta {\cal D}\bar{V}_\eta  { Z}_\eta[V_\eta,\bar{V}_\eta]
e^{-iV_{\eta}\bar{\rho}_{\eta}-i\bar{V}_{\eta}\rho_{\eta}}\,.
\end{align}
Substituting Eq.~(\ref{identity}) into Eq.~(\ref{bosonic_action}), we
obtain the bosonized action
\begin{eqnarray}&&
\label{action_free1}
S_{0\eta}[\rho_\eta,\bar{\rho}_\eta]=
\rho_{\eta}\Pi^{a^{-1}}_\eta\bar{\rho}_\eta -i \ln Z_\eta[\bar{\chi}].
\end{eqnarray}
Here $Z_\eta[\bar{\chi}]\equiv Z_\eta[\chi=0,\bar{\chi}]$
is a partition function (\ref{identity}) of
free fermions,
\begin{equation}
i\ln Z_\eta[\bar{\chi}]=\sum_{n=2}^\infty i^{n+1}
\bar{\chi_\eta}^n{\cal S}_{n,\eta}/n!\,,
\end{equation}
subject to the external quantum field
\begin{equation}
\label{eq_chi}
\bar{\chi}_\eta=\Pi_\eta^{a^{-1}}\bar{\rho}_\eta\,.
\end{equation}
The combined action of left- and right-moving electrons is simply
given by a sum of the corresponding chiral actions,
\begin{equation}
\label{action_free}
S_0[\rho,\bar{\rho}]=\sum_\eta S_{0\eta}[\rho_\eta,\bar{\rho}_\eta]\,.
\end{equation}

\begin{figure}
\includegraphics[width=0.8\columnwidth,angle=0]{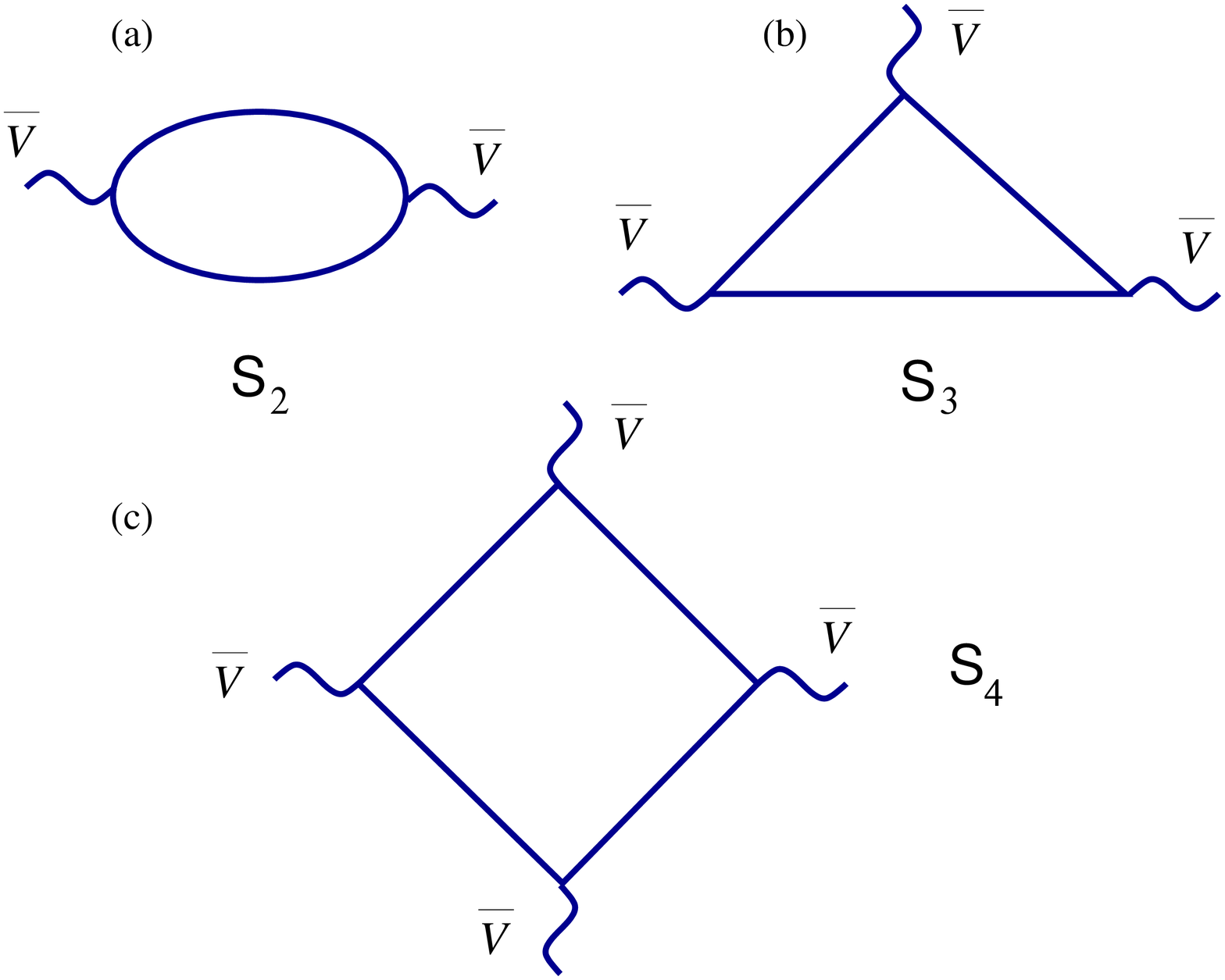}
\caption{Vacuum loops for free fermions in an external field
  $\bar{V}$. At equilibrium only ${\cal S}_2$ is non-zero,
  according to the Dzyaloshinskii-Larkin theorem. Away from
  equilibrium, all vertices appear. For details see Sec.~\ref{sec_DL}.
}
\label{DL}
\end{figure}

Thus we have described a  system of non-equilibrium free fermions
by  a bosonic theory, Eq.(\ref{action_free}).
In this approach  information on the non-equilibrium state of the system
is encoded  in the vertices (${\cal S}_{n\eta}$), schematically depicted in
Fig.\ref{DL}. In Sec.~\ref{sec_DL} we discuss the status and
implications  of the Dzyaloshinskii-Larkin theorem concerning these
vertices.

\subsection{Dzyaloshinskii-Larkin theorem}
\label{sec_DL}

The appearance of higher-order ($n>2$)
fermionic vertices may seem to contradict
the Dzyaloshinskii-Larkin theorem\cite{DzLar:73}.
The latter states that diagrams containing closed loops with more than
two fermionic lines vanish.
Although the theorem was formulated for the equilibrium case,
its proof, given in Ref.~\onlinecite{DzLar:73},
ostensibly relies solely  on  particle conservation.
Since the latter remains  valid out of equilibrium,
one might expect the theorem to hold under non-equilibrium conditions
as well. To understand why  Dzyaloshinskii-Larkin  theorem
is in fact restricted to the equilibrium case only,
and what its implications for a non-equilibrium situation are,
we carefully re-examine the arguments of Ref.~\onlinecite{DzLar:73}.

One starts with the continuity equation for the chiral current
and density operators,
\begin{eqnarray}
\label{cont_operatorial}
\omega{\rho}_{\eta} -
\eta q{j}_{\eta}=0\,.
\end{eqnarray}
Since within the LL model these operators are related
to each other through
$j_{\eta}=\eta v \rho_{\eta}$, the continuity equation can be
rewritten in terms of the density field only,
\begin{eqnarray}
\label{cont_density}
(\omega - \eta vq) {\rho}_{\eta} = 0\,.
\end{eqnarray}
As a consequence, correlation functions of densities satisfy
\begin{equation}
\label{DL-correlators}
(\omega_i - \eta vq_i)\langle{\rho}_\eta(\omega_1,q_1)
{\rho}_\eta(\omega_2,q_2)\dots{\rho}_\eta(\omega_n,q_n)\rangle=0
\end{equation}
for any $i=1,\ldots,n$.
Therefore, the irreducible density correlation functions
${\cal  S}_{n\eta}(\omega_1,q_1;\omega_2,q_2;\dots \omega_n,q_n)$ with
$n>2$ should be zero everywhere, except possibly for the mass shell
with respect to all arguments\cite{note-DL},
\begin{eqnarray}
\label{mass-shell}
&& {\cal S}_{n\eta}(\omega_1,q_1;\omega_2,q_2;\dots \omega_n,q_n) =
\nonumber\\
&& = \delta(\omega_1 - \eta vq_1) \delta(\omega_2-\eta vq_2)\ldots
\delta(\omega_n-\eta vq_n) \nonumber \\
&&\times {\cal S}_{\eta}(\omega_1,\omega_2, \dots, \omega_n)
\delta(\omega_1+\ldots + \omega_n).
\end{eqnarray}
In the case $n=2$ the argument is not applicable in view of the
Schwinger anomaly, yielding the first term in the exponent on the
r.h.s. of Eq.~(\ref{identity}).
When translated into the coordinate-time space, the
mass-shell condition (\ref{mass-shell}) implies that the correlation
function depends in fact only on the world line to which each of the
points $(t_i,x_i)$ belongs but not on the position of this point on
the line:
\begin{eqnarray}
\label{mass-shell-xt}
{\cal S}_{n\eta}(t_1,x_1;\ldots;t_n,x_n) & \equiv &
\langle\langle{\rho}_\eta(t_1,x_1)\ldots {\rho}_\eta(t_n,x_n)\rangle\rangle
\nonumber \\
& = & \langle\langle{\rho}_\eta(0,\xi_1)\ldots
{\rho}_\eta(0,\xi_n)\rangle\rangle, \nonumber \\
&&
\end{eqnarray}
where $\xi_i=x-\eta v t_i$.
In the Keldysh formalism language, the only non-zero irreducible
density correlation
function in any order $n>2$ arises when one considers the correlator
with all $n$ fields being the classical components $\rho$. (This
follows from the fact that the operators $\rho$ commute to a c-number.)
These correlation functions are the noise cumulants in the system.

The behavior of the correlation functions  ${\cal S}_{n\eta}$ on
the ``light cone'' (\ref{mass-shell})  can not be
determined from particle conservation law and requires an additional
calculation. While at equilibrium  all  ${\cal S}_{n\eta}$ with $n>2$
do vanish (which reconciles our theory with the Larkin-Dzyaloshinskii
theorem), out of equilibrium they are in general non-zero. We consider
this general situation in Sec.~\ref{sec_functional_det} where we show
that the bosonized action can be presented in a compact form of a
functional determinant.

\subsection{Bosonized action as functional determinant}
\label{sec_functional_det}

As we have shown, the bosonic action, Eq.~(\ref{action_free}), is
expressed through
the  partition function $Z[V,\bar{V}]$ of free fermions in
an external field $V(x,t)$ defined on the Keldysh contour.
In one dimension the partition function can be cast in a relatively
simple form.
To achieve this, we first present the partition function
\begin{equation}
\label{eq_trace1}
Z[V,\bar{V}]=
\tr\{\rho_F S_c\}\,.
\end{equation}
Here $S_c$ is an evolution operator along Keldysh contour,
\begin{eqnarray}
\label{eq_trace2}
Z[V,\bar{V}]&=&\lim_{t \rightarrow \infty}
\tr\{\rho_F
e^{-iH[V_+(-t)]\Delta t} e^{-iH[V_+(-t+\Delta t)]\Delta t} \nonumber \\
& \times & \ldots \times
e^{-iH[V_+(t)]\Delta t}e^{iH[V_-(t) ]\Delta t}
e^{iH[V_-(t-\Delta t) ]\Delta t}
 \nonumber \\
& \times & \ldots \times
e^{iH[V_-(-t) ]\Delta t}
\}\,,
\end{eqnarray}
and the trace is taken over the  many-body fermionic Fock space.
Equation (\ref{eq_trace2}) can be further simplified
by means of the following identity\cite{Klich}
\begin{eqnarray}&&
\label{eq_Klich}
\tr\{e^{H_1}e^{H_2}\dots e^{H_N}\}=
\det(1+e^{h_1}e^{h_2}\dots e^{h_N})\,.
\end{eqnarray}
Here $h_n$ is a matrix in the single particle  Hilbert space, and
\begin{equation}
H_n=\sum_{i,j}h_n^{i,j}a^\dagger_ia_j
\end{equation}
is the corresponding  operator quadratic in fermionic
creation/annihilation operators ($a^\dagger,a$).
The trace in the l.h.s.  of Eq.(\ref{eq_Klich}) is taken in  the
many-body Fock space, while the determinant on the r.h.s.
is taken in the single-particle space.

Applying Eq.(\ref{eq_Klich}) in the continuum limit, we express the
partition function  in the following form
\begin{equation}
\label{eq_det}
Z_\eta[V_\eta,\bar{V}_\eta]=\det[1-{n}_\eta+{n}_\eta
U_{+,\eta}^{-1} U_{-,\eta}]\,.
\end{equation}
Here
\begin{eqnarray}&&
U_{+,\eta}(t)=\mathrm{Texp}\left(-i\int_0^tdt{h}_{+,\eta}\right) \, ,
\nonumber  \\&&
U_{-,\eta}^{-1}(t)=\mathrm{\tilde{T}exp}\left(i\int_0^tdt{h}_{-,\eta}\right)\,
\end{eqnarray}
are evolution operators  that correspond to the
single-particle Hamiltonians
\begin{eqnarray}&&
{h}_{+,\eta}=-i\eta v\frac{\partial}{\partial x}+V_+(x,t)\,,
\nonumber  \\&&
{h}_{-,\eta}=-i\eta v\frac{\partial}{\partial x}+V_-(x,t)\,.
\end{eqnarray}
Thus the many-body problem of summing all vacuum loops has been reduced
to a calculation of a functional determinant
of an operator in a single-particle Hilbert space.
To simplify it further, we analyze the properties of the
evolution operator $U$ in one dimension.
Its action on a wave function $\psi(x)$ can be described as
\begin{eqnarray}&&
\label{Schrodinger_eq1}
\psi(x,t)=\mathrm{Texp}\left(-i\int_0^tdt{h}_{+}\right)\psi(x,0)\,,
\end{eqnarray}
where $\psi(x,0) \equiv \psi(x)$.
One can easily show that the resulting wave function $\psi(x,t)$
satisfies the Schr{\"o}dinger equation
\begin{equation}
\label{Schrodinger_eq2}
i\frac{\partial}{\partial t}\psi(x,t)=h_+\psi(x,t)\,.
\end{equation}
Solving Eq.(\ref{Schrodinger_eq2}) explicitly one finds
\begin{equation}
\psi(x,t)=\psi(x-\eta vt,0)e^{-i\int_0^td\tau V_+(x+\eta v(\tau-t),\tau)}\,.
\end{equation}
Therefore, the action on a wave function of the evolution operator
forward and backward in time results in the phase factor
\begin{equation}
(U^{-1}_-U_+\psi)(x)=\psi(x)e^{-i\int_0^td\tau(V_+-V_-)(x+v\tau,\tau)}\,.
\end{equation}

Consequently, the partition function of the 1D fermions
can be cast as\cite{anomalous}
\begin{equation}
\label{eq_det2}
Z_\eta[V_\eta,\bar{V}_\eta]=
e^{-iV_{\eta}\Pi^{a}_\eta \bar{V}_{\eta}}
\Delta_\eta[\delta_\eta(t)]\,,
\end{equation}
where we introduced a determinant
\begin{equation}
\label{determinant_definition}
\Delta_\eta[\delta_\eta(t)]=
\det[1
  + (e^{-i{\delta_\eta}}-1){n}_\eta]\,,
\end{equation}
and
\begin{equation}
\label{phase}
\delta_\eta(t)=\sqrt{2}\int_{-\infty}^{\infty}d\tau\bar{V}_\eta(\eta
v(\tau+t),\tau)
\end{equation}
is the scattering phase accumulated by an electron moving
along a ``light-cone'' trajectory. Thus, according to
Eq.~(\ref{eq_det2}) the problem of
summing up the vacuum loops is
reduced to evaluation of a one-dimensional functional determinant
(\ref{determinant_definition}).

The determinant (\ref{determinant_definition}) is defined by the
function $\delta_\eta(t)$ in the time space and $n_\eta(\epsilon)$ in
the energy space, with $\epsilon$ and $t$ understood as canonically
conjugate variables. It belongs to the class of Fredholm determinants.
For a specific case (that will be particularly important for us below)
when  $\delta_\eta(t)$ is different from zero
within a limited interval of time only, the determinant acquires the
Toeplitz form. Such  determinants have been considered in the context
of counting statistics \cite{Levitov-noise,counting-statistics}; see a
more detailed
discussion in Sec.~\ref{sec:Green_functions}.
It is also worth mentioning that there is a vast literature on
the connection of Fredholm determinants to quantum integrable models,
classical integrable differential equations (with soliton solutions),
and free-fermion problems; we refer the reader to
Refs.~\cite{Jimbo80,Izergin98,bettelheim06,Zvonarev} and
references therein.

At equilibrium the Taylor expansion of $\ln Z$ in $\delta$ terminates
at the second order  (${\cal S}_n=0$ for $n>2$),
in agreement with Dzyaloshinskii-Larkin theorem, Ref.~\onlinecite{DzLar:73}.
In that case the action  (\ref{action_free}) is quadratic,
reproducing  the standard  LL model.
Away from thermal equilibrium, high-order
density correlations are finite \cite{Levitov-noise}.
For this reason, we obtain a non-Gaussian bosonized theory,
despite the fact that the Hamiltonian (\ref{Hamiltonian_free_bosons}) is
quadratic. The higher-order terms ${\cal S}_n$ with $n>2$
appear in the bosonic action due to a
non-diagonal structure of the density matrix in the bosonic Fock
space, which leads to a breakdown of  Wick theorem for the bosonic
fields.

\subsection{Green functions}
\label{sec:Green_functions}

We have thus shown that non-interacting fermions can be equivalently
described by the bosonic theory with the action given by
Eqs.~(\ref{action_free}), (\ref{action_free1}), (\ref{eq_det2}). We
apply now this formalism to calculate the free-fermion Green functions
(GFs),
\begin{eqnarray}&&
\label{eq_green_function}
G^<_{\eta}(x_1,t_1;x_2,t_2)=i\langle
\psi_\eta^\dagger(x_2,t_2)\psi_\eta(x_1,t_1)\rangle,
\nonumber \\&&
G^>_{\eta}(x_1,t_1;x_2,t_2)=-i\langle
\psi_\eta(x_1,t_1)\psi^\dagger_\eta(x_2,t_2)\rangle\,.
\end{eqnarray}
At equilibrium these GF's are related to the advanced and retarded
GFs via
\begin{eqnarray}&&
\label{G>}
G_\eta^>(x,\epsilon)=[G^r_\eta(x,\epsilon)-G^a_\eta(x,\epsilon)]
(1-n_\eta(\epsilon))\,,
\nonumber \\&&
G_\eta^<(x,\epsilon)=-[G^r_\eta(x,\epsilon)-G^a_\eta(x,\epsilon)]
n_\eta(\epsilon)\,.
\end{eqnarray}
For free fermions, Eq.~(\ref{G>}) is  valid for an arbitrary
distribution function $n_\eta$ determining the filling of
single-particle states.

Due to Galilean invariance  the GFs depend only on
$\tau_\eta=t_1-t_2-\eta (x_1-x_2)/v$,
so we may set $x_1=x_2=vt_2=0$ in the argument of GF.
Using Eqs.~(\ref{fermi_operator2}), (\ref{eq_green_function}), we obtain
\begin{align}&&
\label{Green_function}
G^>_{0,\eta}(\tau_\eta)=-\frac{i\Lambda}{2\pi v}\langle T_K
e^{i{\phi}_{\eta,-}(0,\tau_\eta)}e^{-i{\phi}_{\eta,+}(0,0)}\rangle\,,
\end{align}
and a similar result for the function $G^<_{0,\eta}$.
At thermal equilibrium  $G^\gtrless_{0,\eta}$
can be readily calculated. A standard calculation
(presented  for completeness in  Appendix \ref{appendix1})
yields
\begin{equation}
\label{Green_function_equilibrium}
G^\gtrless_{\eta}(\tau_\eta)=\mp\frac{i\Lambda}{2v}
\frac{T\tau_\eta}{\sinh \pi T\tau_\eta }\frac{1}{1\pm i\Lambda\tau_\eta}\,.
\end{equation}

Away from equilibrium the calculation of GF's,
rather simple  within a fermionic framework,
turns out to be quite complicated within a bosonic one.
Nevertheless, this effort pays off, since the bosonic formalism will
later allow us to extend the analysis to the interacting case.

Within the bosonic description  the GF
can be represented  as a functional integral
over the density fields. Since calculations of  $G_{0,\eta}^>$ and $G_{0,\eta}^<$
are quite similar to each other we focus here on
\begin{eqnarray}
\label{Green_function2}
G^>_{0,\eta}(\tau_\eta)
&=& -\frac{i\Lambda}{2\pi v}\int {\cal D}\rho{\cal D}\bar{\rho}
e^{iS[\rho,\bar{\rho}]}\nonumber \\ &\times &
e^{{i\over\sqrt{2}}[\phi(0,\tau_\eta)-\phi(0,0)-\bar{\phi}(0,\tau_\eta)-
\bar{\phi}(0,0)]}.
\end{eqnarray}

In a generic non-equilibrium situation the bosonic action,
Eq.~(\ref{action_free}), contains terms of all orders with no small
parameter; the idea to proceed
analytically in a controlled  manner may seem hopeless. This,
however,  is not the case: non-equilibrium bosonization is an
efficient framework in which the functional integration can be
performed exactly. Indeed,
$Z_\eta$ in Eq.~(\ref{action_free1})
depends only on the quantum component $\bar{\rho}$, so that
the   action, Eq.~(\ref{action_free}), is linear with respect  to the
classical component $\rho$ of the density field. Hence the
integration with respect to $\rho$ can be performed exactly
\begin{eqnarray}
\label{Green_function3}
G^>_{0,\eta}(\tau)&=&-\frac{i\Lambda}{2\pi v}\int {\cal D}\bar{\rho}
Z_\eta[\bar{\chi}_\eta]
\delta(\partial_t\bar{\rho}+\eta v\partial_x\bar{\rho}-j)
\nonumber \\&\times &
e^{{-i\over\sqrt{2}}[\bar{\phi}(0,\tau)+\bar{\phi}(0,0)]}\,,
\end{eqnarray}
where the source term is
\begin{equation}
\label{source_term}
j(x,t)= \delta(x)[\delta(t-\tau)-\delta(t)]/\sqrt{2}.
\end{equation}
Resolving the $\delta$-function, we obtain an
equation that determines the quantum component of the density field,
\begin{equation}
\label{d1}
\partial_t\bar{\rho}_\eta+\eta v\partial_{x}\bar{\rho}_\eta= j(x,t)\,.
\end{equation}
According to the structure of the first term in the action
(\ref{action_free1}), we should look for the advanced solution of
Eq.~(\ref{d1}) which is zero at times larger than those at which the
source $j(x,t)$ acts. In other words, in the asymptotic regions
$|x|>L/2$  the solution $\bar{\rho}(x,t)$ should contain incoming
waves only.  
Solving Eq.~(\ref{d1}) with this asymptotic conditions, 
we find the quantum density component
\begin{eqnarray}
\label{eq_rho_saddle_point}
\bar{\rho}_{\eta}(x,t)&=&\frac{\theta(-\eta x)}{\sqrt{2}}
\bigg[\delta(x-\eta v t)-\delta(x-\eta v(t-\tau)) \bigg]\,.
\nonumber
\\&&
\end{eqnarray}
To find the Green function, we need to evaluate the factors
multiplying the delta-function in Eq.~(\ref{Green_function3}),
subjected to the $\delta$-function constraint. 
The most non-trivial factor (which carries the information
about the distribution function) is $Z_\eta[\bar{\chi}_\eta]$, where
$\bar{\chi}_\eta$ is related to $\bar{\rho}_\eta$ via
Eq.~(\ref{eq_chi}). According to Eq.~(\ref{eq_det2}),
  $Z_\eta[\bar{\chi}_\eta]$ is expressed as a functional determinant
of the form  (\ref{determinant_definition}). We thus obtain
\begin{equation}
\label{a1}
G_{0,\eta}^\gtrless(\tau)=-\frac{1}{2\pi v}\frac{1}{\tau\mp i/\Lambda}
\overline{\Delta}_\eta[\delta_\eta(t)]\:.
\end{equation}
Here we have denoted by $\overline{\Delta}_\eta$ the determinant
normalized to its value for zero-temperature equilibrium distribution,
see Appendix A. It is convenient to use this definition since the
determinant ${\Delta}_\eta$ requires in fact an ultraviolet
regularization. On the other hand, the normalized determinant
$\overline{\Delta}_\eta$ (which is equal to unity for the equilibrium,
$T=0$ case) is uniquely defined. The prefactor in Eq.~(\ref{a1}) that
does not depend on the distribution function is immediately determined
from the equilibrium result.

According to Eqs.(\ref{eq_det2}), (\ref{phase}), the mass-shell
nature of $S_{\eta n}$ implies that $Z_\eta[\bar{\chi}_\eta]$
depends only on the world-line integral
\begin{equation}
\label{d1a}
\delta_\eta(t)=\sqrt{2} \int_{-\infty}^\infty
d\tilde{t}\: \bar{\chi}_\eta(
\eta v\tilde{t},
\tilde{t}-t).
\end{equation}
Using Eq.~(\ref{eq_chi}), we find an explicit
solution for the ``counting field'' $\bar{\chi}_\eta$,
\begin{eqnarray}
\label{chi}
\bar{\chi}_\eta(x,t)=2\pi[v\bar{\rho}_\eta(x,t)+\eta\int_0^xd\tilde{x}
\bar{\rho}_\eta(\tilde{x},t)].
\end{eqnarray}
Next we calculate the value of $\delta(t)$ for our non-interacting problem.
Substitution of Eq.~(\ref{chi}) into Eq.~(\ref{d1a}) allows us to cast
the result for the phases $\delta_\eta(t)$ into the following form:
\begin{equation}
\label{a7a}
\delta_\eta(t)=
-2\pi\sqrt{2}\:
\eta\lim_{\tilde{t}\rightarrow-\infty}\int_0^{\eta v(\tilde{t}+t)}d\tilde{x}\:
\bar{\rho}_\eta(\tilde{x},\tilde{t})\,.
\end{equation}

For the free-fermion problem the phase
$\delta_\eta(t)=\lambda\omega_\tau(t,0)$
where
\begin{equation}
w_\tau(t,\tilde{t})=\theta(\tilde{t}-t)-\theta(\tilde{t}-t-\tau)
\end{equation}
is a ``window function'' and
$\lambda=2\pi$. Thus, $Z_\eta[\bar{\chi}_\eta] =
{\overline{\Delta}}_{\eta\tau}(2\pi)$, where
${\overline{\Delta}}_{\eta\tau}(\lambda)$ is the
determinant (\ref{determinant_definition}) (normalized to its $T=0$
value) for a rectangular pulse.
\begin{equation}
\label{identity-green-function}
G_{0,\eta}^\gtrless(\tau)=-\frac{1}{2\pi
  v}\frac{{\overline{\Delta}}_{\eta\tau}(2\pi)}{\tau\mp i/\Lambda}
\:.
\end{equation}

Determinants of the type (\ref{determinant_definition})
have appeared  in a theory of counting statistics
\cite{Levitov-noise,counting-statistics}.
Specifically, the generating function of current
fluctuations
$\kappa(\lambda)=\sum_{n=-\infty}^\infty e^{i n\lambda}p_n$ (where $p_n$
is the  probability of  $n$ electrons being transferred through the
system in a given time window $\tau$) has the same structure as
$\Delta_{\eta\tau}(\lambda)$. Taylor expansion of
$\ln\kappa(\lambda)$ around $\lambda=0$ defines cumulants of current
fluctuations.

\begin{figure}
\includegraphics[width=0.8\columnwidth,angle=0]{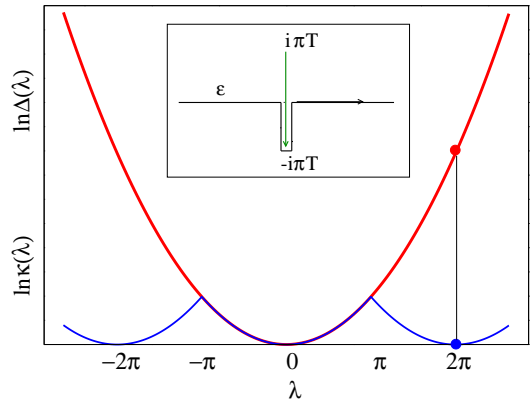}
\caption{Analytic ${\overline{\Delta}}_{\eta\tau} (\lambda)$ vs. periodic
  $\kappa(\lambda)$ continuation of the functional determinant. The value of
  ${\overline{\Delta}}_{\eta\tau}$ at  $\lambda=2\pi$ determines the
  free-electron
  GF, while  $\ln\kappa(2\pi)=0$ in view of charge quantization.
As an example, the equilibrium case is shown.
 {\it
    Inset:} contour of integration for the quasiclassical limit,
  Eq.~(\ref{a5}),  of
  ${\overline{\Delta}}_{\eta\tau}(\lambda)$ is deformed, since a
  singularity of the integrand
  crosses the real axis at $\lambda=\pi$.}
\label{fig5a}
\end{figure}

According to its definition, $\kappa(\lambda)$ is $2\pi$-periodic,
which is a manifestation of  charge quantization that should show up
in measurements of the transfered electric charge
\cite{Levitov-noise,counting-statistics,Klich,Reznikov,Glattli,Devoret,Prober}. Thus,  
$\kappa(2\pi)= 1$ is trivial. On the other hand, we have found
that the free electron GF is determined by the  non-trivial value of
the functional determinant exactly at  $\lambda=2\pi$. A resolution
of this apparent paradox is as follows: the determinant
$\Delta_{\eta\tau}(\lambda)$ should be understood as an analytic
function of $\lambda$ increasing from 0 to $2\pi$. On the other
hand, $\kappa(\lambda)$ is non-analytic at the branching points
$\lambda=\pm\pi, \pm 3\pi,\ldots$.
To demonstrate this, it is instructive to consider the equilibrium case
that is treated in Appendix \ref{appendix1}.
Then the expansion of
$\ln\Delta_{\eta\tau}(\lambda)$ in $\lambda$ is restricted to the
$\lambda^2$ term (since RPA is exact). It is easy to check that the
$\lambda=2\pi$ point on this parabolic dependence correctly
reproduces the fermion GF via Eqs.~(\ref{a1}),
(\ref{determinant_definition}).
As to the counting statistics $\ln\kappa(\lambda)$, it
is quadratic only in the interval $[-\pi,\pi]$ and is periodically
continued beyond this interval, see Fig.~\ref{fig5a}.

The difference in the analytical properties of $\kappa(\lambda)$ and
$\Delta_{\eta\tau}(\lambda)$ becomes especially transparent if one studies
the semiclassical (long-$\tau$) limit,
\begin{equation}
\label{a5}
\ln \bar{\Delta}_{\eta\tau}(\lambda)=\frac{\tau}{2\pi}
\int_{-\infty}^{\infty} d\epsilon\bigg\{
\ln[1+(e^{-i\lambda}-1) n_\eta(\epsilon)]+i\lambda\theta(-\epsilon)
\bigg\}\,.
\end{equation}
For small positive $\lambda$ the singularity of the integrand closest to the
real axis is located at $\epsilon =  i(\pi-\lambda) T$, i.e.
near $\epsilon = i\pi T$. As $\lambda$
increases, the singularity moves towards the real axis, crosses it
at $\lambda=\pi$ and finally approaches $\epsilon=-i\pi T$ as
$\lambda\to2\pi$ (see  inset of Fig.~\ref{fig5a}). The integral for
$\ln\kappa(\lambda)$ is taken along the real axis, resulting in
non-analyticity at $\lambda=\pi$ and in zero value at
$\lambda=2\pi$. On the other hand, the contour of energy  integration for
$\ln\bar{\Delta}_{\eta\tau}(\lambda)$ with $\lambda>\pi$
is deformed in the complex plane to preserve analyticity,
as shown in Fig. \ref{fig5a}.
Specifically, the contour  consists of the integration along the real
axis a part along the branch cut on the imaginary axis.
The integration along the real axis yields
\begin{equation}
\int_{-\infty}^{\infty}d\epsilon\bigg[
\ln\left(\frac{e^\frac{\epsilon}{T}+e^{-i\lambda}}
{1+e^\frac{\epsilon}{T}}\right)+i\lambda\theta(-\epsilon)\bigg]
=
-\frac{T\tilde{\lambda}^2}{2}\,,
\end{equation}
where $N\equiv [\lambda/2\pi]$,
$\lambda=\tilde{\lambda}+2\pi N$.
The integration along the branch cut
of the logarithm yields $-(T/2)[(2\pi N)^2+4\pi N \tilde{\lambda}]$,
resulting in the long-$\tau$ asymptotics
\begin{equation}
\ln{\overline{\Delta}}_{\eta\tau} =-\tau T\lambda^2/4\pi.
\end{equation}
Substituting this in Eq.~(\ref{identity-green-function}), we correctly
reproduce the long-time asymptotics of the Green function $G^>_0$
at equilibrium, Eq.~(\ref{Green_function_equilibrium}).

Let us now turn to the non-equilibrium situation and
consider the double step function
\begin{equation}
n_\eta(\epsilon)=a_\eta
n_0(\epsilon_-)+ (1-a_\eta)n_0(\epsilon_+) \,,
\label{double-step}
\end{equation}
where $n_0(\epsilon)=\theta(-\epsilon)$ is the zero-temperature
Fermi-Dirac distribution function, $\epsilon_\pm =
\epsilon-\mu \pm V/2$,  and $0<a_\eta <1$.
The value of $\mu$ is fixed by demanding
that the total number of electrons is the same as for the equilibrium
distribution $n_0(\epsilon)$ (which we use for normalization), yielding
$\epsilon_-=\epsilon-(1-a)eV$, $\epsilon_+=\epsilon+aeV$.
The distribution function in the time domain
\begin{eqnarray}&&
\label{distribution_function_time_domain}
n_\eta(\tau)=\int_{-\infty}^{\infty} \frac{d\epsilon}{2\pi} e^{-i \epsilon\tau+
0\epsilon }n_\eta(\epsilon)
\end{eqnarray}
can be straightforwardly calculated,
and is given by a sum of oscillating terms
\begin{eqnarray}
\label{distribution_function_time_domain2}
n_\eta(\tau) &=&
(1-a_\eta)e^{ia_\eta eV\tau} n_0(\tau)
\nonumber \\&+&
a_\eta e^{-ieV\tau(1-a_\eta)} n_0(\tau)
\,,
\end{eqnarray}
where
\begin{eqnarray}
n_0(\tau)=\frac{i}{2\pi}\frac{1}{\tau+i0}
\end{eqnarray}
is  the $T=0$ Fermi-Dirac distribution function in time representation.

On the other hand, we can find
the time dependence of the fermionic distribution function by using
our non-equilibrium bosonization approach, leading to
the identity (\ref{identity-green-function}). In the long-time limit
we need to evaluate the integral (\ref{a5}), yielding
\begin{equation}
\ln {\overline{\Delta}}_{\eta\tau}(\lambda)
\simeq
\frac{eV\tau}{2\pi}\bigg(\ln\left(1-a_\eta+a_\eta e^{-i\lambda}\right)+
ai\lambda\!\bigg)\,.
\end{equation}
Analytically continuing in $\lambda$  we get
\begin{eqnarray}
\ln {\overline{\Delta}}_{\eta\tau}(2\pi)\simeq
ieV\tau
\left\{\!\!\begin{array}{l}
a_\eta-1  \,, \,\,\, a_\eta>1/2
\\
\\
a_\eta\, , \hspace{0.5cm}  a_\eta<1/2\,,
\end{array}
\right.
\end{eqnarray}
which reproduces the long-time time limit of the Green function of
free fermions with  the distribution function
(\ref{distribution_function_time_domain2}).  We have just demonstrated
how the identity   (\ref{identity-green-function})  works for a
double-step non-equilibrium distribution.

Equation (\ref{identity-green-function})
is a remarkable identity, as it connects two
seemingly unrelated objects: the distribution function of free fermions
and a Fredholm determinant of the counting operator.
The value of $\lambda=2\pi$ appearing in the bosonic representation
of the free-fermion GF $G_{0,\eta}(\tau)$ has a clear
physical meaning: a fermion is a $2\pi$-soliton in the bosonic
formalism.

\vspace{0.5cm}

\section{Fermi Edge Singularity}
\label{sec_FES}
 A natural question to ask is whether  values of
$\Delta_{\eta\tau}(\lambda)$ away from $\lambda=2\pi$ are physically
important. To see that this is indeed the case, consider  the Fermi
edge singularity (FES) problem. In this problem, an electron excited
into the conduction band, leaves behind a localized hole, resulting
in an $s$-wave scattering phase shift, $\delta_0$, of the
conducting electrons \cite{Nozieres}. In the mesoscopic
context\cite{MatveevLarkin},
the FES manifests itself in resonant tunneling experiments \cite{Geim}.
On a formal level it is described by the following Hamiltonian
\begin{eqnarray}
H=\sum_k\epsilon_k a_k^\dagger a_k+E_0b^\dagger
b+\sum_{k,k'}V_{k,k'}a^\dagger_k a_{k'}bb^\dagger\,.
\end{eqnarray}
While in the FES problem there is no interaction between electrons
in the conducting band,
it has many features characteristic of genuine many-body physics.
Historically, the FES problem was first solved by an exact summation of an
infinite diagrammatic series \cite{Nozieres}.
Despite the fact that conventional experimental realizations of FES are
three-dimensional, the problem can be reduced (due to the local
character of the interaction with the core hole) to that of
one-dimensional chiral fermions. For this reason,  bosonization
technique can be effectively applied, leading
to  an alternative and very elegant solution \cite{Schotte}.

Away from equilibrium, the FES  has been addressed in
Ref.~\onlinecite{Abanin} where the canonical (fermionic) FES theory
was combined with the scattering matrix approach.
Below, we apply the non-equilibrium bosonization technique
to the same problem.

As mentioned above,
the FES problem is effectively described by chiral 1D electrons
interacting with a core hole that is instantly ``switched on''.
As was shown in \cite{Schotte}, taking into account the core hole in
the bosonization approach amounts to  replacement of
$e^{i{\phi}}$ by $e^{i(1-\delta_0/\pi){\phi}}$ in the boson
representation of the fermionic operator.
Using Eqs.~(\ref{fermi_operator2}), (\ref{eq_green_function}), one gets
\begin{align}&&
\label{Green_function_FES}
G^>(\tau)=-\frac{i\Lambda}{2\pi v}\langle T_K
e^{i\left(1-\frac{\delta_0}{\pi}\right){\phi}_{-}(0,\tau)}
e^{-i\left(1-\frac{\delta_0}{\pi}\right){\phi}_{+}(0,0)}\rangle\,
\end{align}
and similarly for the function $G^<$.
Within our non-equilibrium formalism, this implies a replacement
$j\rightarrow(1-\delta_0/\pi)j$ in Eq.~(\ref{d1}). Performing the derivation as
in the free-fermion case, we thus obtain the non-equilibrium FES
GF for electrons with an arbitrary distribution
$n(\epsilon)$,
\begin{equation}
\label{b1}
G^\gtrless(\tau)=\mp i\Lambda{\overline{\Delta}}_{\tau}(2\pi-2\delta_0) /
2\pi v (1\pm i\Lambda\tau)^{(1-\delta_0/\pi)^2}.
\end{equation}
At equilibrium Eq.~(\ref{b1}) can be further simplified (see Appendix A),
\begin{equation}
\label{b1a}
G^\gtrless(\tau)= \left(\frac{\pi T\tau}{
\sinh \pi T \tau}\right)^{(1-\frac{\delta_0}{\pi})^2}
\frac{\mp i\Lambda}{
2\pi v (1\pm i\Lambda\tau)^{(1-\frac{\delta_0}{\pi})^2}}\,,
\end{equation}
reproducing the known results \cite{Nozieres,Schotte}.

For a double-step distribution, Eq.~(\ref{double-step}),
the long-time limit is obtained as
\begin{equation}
\label{eq_79}
\Delta_\tau(2\pi-2\delta_0) \simeq e^{-\tau/2\tau_\phi(2\delta_0)}\,,
\end{equation}
where the dephasing rate $\tau_\phi^{-1}$ is given by
\begin{equation}
\label{eq_80}
\tau^{-1}_\phi(\lambda)=
-\frac{eV}{2\pi}\ln\left(1-4a(1-a)\sin^2\frac{\lambda}{2}\right)\,.
\end{equation}
In the energy representation $\tau_\phi^{-1}$ determines the broadening
of the split  FES singularities.
The same result for the broadening of FES has been obtained by Abanin
and Levitov in  Ref.~\onlinecite{Abanin} within the fermionic
framework.
It is instructive to compare their result  with our
analysis.  In  the bosonization technique we have expressed
the GF of the FES problem in terms of a functional determinant (\ref{b1}).
On the other hand, within the fermionic approach \cite{Abanin}
the GF splits  into a product of an open line $L(\tau)$
(i.e. single particle Green function of fermions
in the presence of external time dependent field)
and closed loop $e^C$  (i.e. vacuum loops of fermions in an external
field),
\begin{equation}
\label{eq_fes}
G^\gtrless(\tau)=L^\gtrless(\tau)e^C\,,
\end{equation}
with the closed-loop part given by
\begin{equation}
\label{closed_loops}
e^C=\Delta_\tau(-2\delta_0)\,.
\end{equation}
This  representation of the Green function is similar to the
functional bosonization approach \cite{Fogedby:76,LeeChen:88,Naon,Yur},
that employs both fermionic and bosonic variables.
While functional and full bosonization approaches yield equivalent
results, this equivalence is highly non-trivial.
Indeed, comparing Eq.~(\ref{eq_fes}) with (\ref{b1})
and employing Eq.~(\ref{closed_loops}),
we  establish the identity
\begin{eqnarray}
\label{idenity_explicit}
\mp\frac{i\Lambda}{2\pi v}
 \left(\frac{1\mp i\tau\Lambda}{1\pm i\Lambda\tau}\right)^{(1-\delta_0/\pi)^2}
\Delta_\tau(2\pi-2\delta_0)  && \nonumber \\
 =L^\gtrless(\tau){\Delta}_\tau(-2\delta_0)\, &&
\end{eqnarray}
relating the functional determinants
$\overline{\Delta}_\tau(2\pi-2\delta_0)$ and
$\overline{\Delta}_\tau(-2\delta_0)$
through the single-particle Green function $L(\tau)$.

Since ${n}_\eta$ is diagonal in energy space, while
${\delta_\eta}$ is diagonal in time space, they do not commute,
making the determinant non-trivial.
It is worth noting  that  the
functional determinants  ${\overline{\Delta}}_\tau(\lambda)$
for $|\lambda| <\pi$ have been efficiently studied by
numerical means \cite{Nazarov,Marquardt}.
The  identity (\ref{idenity_explicit}) can be useful for the numerical
evaluation of  ${\overline{\Delta}}_\tau(\lambda)$ at larger values of
$\lambda$.

\section{ interacting electrons}
\label{sec_interaction}

So far we have been dealing with non-interacting electrons.
Now we focus on the main subject of this work:
bosonization of interacting fermions,  both for spinless and
for spinful cases. We begin by showing in Sec.~\ref{sec_Keldysh_action}
how the interaction can be incorporated
into the non-equilibrium bosonization scheme developed above.

\subsection{Keldysh action}
\label{sec_Keldysh_action}

\begin{figure}
\includegraphics[width=\columnwidth,angle=0]{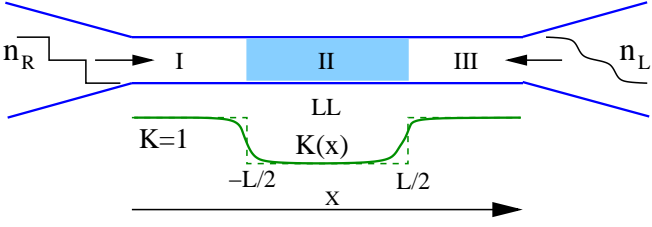}
\caption{Schematic view of a LL conductor connected to leads
with two different incoming fermionic distributions. The LL
interaction parameter $K(x)$ is also shown; the dashed line
corresponds to its sharp variation at the boundaries.}
\label{fig1}
\end{figure}

For the problem of spinless interacting fermions  the Hamiltonian reads
\begin{eqnarray}&&
H=H_0+H_{\rm ee}\,,
\label{Ham_int_1}
\end{eqnarray}
where $H_0$ is given by Eq.~(\ref{Hamiltonian_free}) and
\begin{eqnarray}
H_{\rm ee}=\frac{1}{2}\int dx g(x)\left(\rho_L(x)+ \rho_R(x)\right)^2\,,
\label{Ham_int_2}
\end{eqnarray}
where $g(x)$ is a spatially dependent interaction strength. To model
the coupling with non-interacting leads, we will assume that $g(x)$ is
constant within the interacting part of the wire and ``switches off''
near the end points, $x=\pm L/2$, see Fig.~\ref{fig1}. This way of modeling
leads was introduced in Refs.~\onlinecite{Maslov,Ponomarenko,Safi}
to study the conductance of a LL wire; it was also exploited in
Refs.~\onlinecite{Ponomarenko96,Trauzettel04} to analyze the shot noise.
In the Lagrangian formulation, Eqs.~(\ref{Ham_int_1}),
(\ref{Ham_int_1})
 correspond to the action
\begin{eqnarray*}&&
\label{TL}
S[\psi]=S_0[\psi]+S_{\rm ee}[\psi]\,,\nonumber \\&&
S[\psi]=\int_c dt \int dx \sum_{\eta}\bigg[
\psi^\dagger_\eta
i\partial_\eta\psi_\eta
-\frac{g(x)}{2} \rho^2(x,t)\bigg] \,,
\end{eqnarray*}
where $\rho(x)=\rho_{R}(x)+\rho_{L}(x)$.
Decoupling the interaction term via a bosonic field $\varphi$
by means of a Hubbard-Stra\-to\-no\-vich transformation, we  obtain the action
\begin{eqnarray}
\label{TL2}
S[\psi,\varphi]&=&\int_c dt \int dx\bigg[ \sum_{\eta=R,L}\psi^\dagger_\eta
(i\partial_\eta-\varphi)\psi_\eta \nonumber \\
&-&\frac{1}{2}\varphi g^{-1}(x)\varphi\bigg]\,.
\end{eqnarray}
The theory of fermions in an arbitrary field $\varphi(x,t)$
(on the Keldysh contour)
can be bosonized using the results of
Sec.~\ref{sec_free_fermions}. Introducing, as before, notations with
(without) bar for the quantum (classical) components, we obtain the action
\begin{equation}
S[\varphi,\bar{\varphi},\rho,\bar{\rho}]
= S_{0}[\rho,\bar{\rho}] +S_{ee}[\rho,\varphi]\,,
\end{equation}
where  $S_{0}[\rho,\bar{\rho}]$ is the bosonized action of
non-equilibrium free fermions, Eq.(\ref{action_free}) and
\begin{equation}
S_{ee}[\rho,\varphi]
=-\int dtdx[ \varphi \bar{\rho}+\bar{\varphi}\rho
+ \varphi g^{-1}(x)\bar{\varphi}]\,.
\end{equation}
Integrating out the auxiliary
Hubbard-Stratonovich field $\varphi$,
we derive a theory written solely in terms
of density fields,
\begin{eqnarray}&&
\label{TL3}
S[\rho,\bar{\rho}]=S_0[\rho,\bar{\rho}]-
\int dtdx g(x)\rho\bar{\rho}\,.
\end{eqnarray}
Equation (\ref{TL3}) constitutes a bosonic description  for
interacting electrons out of equilibrium.

\subsection{Tunneling spectroscopy of interacting fermions, spinless case}
\label{spinless}

We are now prepared to address  the problem formulated in the
beginning of the paper: an interacting quantum wire out of
equilibrium, Fig.\ref{fig1}. We will first calculate the GFs
at coinciding spatial points, which corresponds to
tunneling spectroscopy measurements. In Sec.~\ref{sec_different_points} we
will generalize this analysis to GFs at different spatial points which
are, in particular, relevant to experiments on LL interferometers.

\subsubsection{Tunneling into the interacting  part of the wire}
\label{sec_tunnel_int}

We consider $G_R^\gtrless(\tau)$ for the
tunneling point ($x=0$) located inside the interacting part of the
wire (region II in Fig.~\ref{fig1});
generalization to tunneling into one of non-interacting leads (regions
I and III  in Fig.~\ref{fig1}) is straightforward and will be
presented in Sec.~\ref{sec_nonint_regions}.

Proceeding in the same way as for the non-interacting case, we come to
a representation of the GF in the form of an integral over the density
fields $\rho$ and $\bar{\rho}$, Eq.~(\ref{Green_function2}).
The only difference as compared to the non-interacting case is that
the bosonic action (\ref{TL3}) now contains also the second term
induced by the interaction. Since this term is linear in the classical
component $\rho_\eta$, we can perform the integration over it
in the same way as we did in the non-interacting case.
As a result, we obtain equations satisfied by the
quantum components $\bar{\rho}_\eta$ of the density fields,
\begin{eqnarray}&&
\label{a7}
\partial_t\bar{\rho}_{R}+\partial_{x}\left
[(v+{g\over 2\pi})\bar{\rho}_{R}+\frac{g}{2\pi}\bar{\rho}_{L}\right]=j\,,
\nonumber \\&&
\partial_t\bar{\rho}_{L}-\partial_{x}\left[(v+\frac{g}{2\pi})\bar{\rho}_{L}+
\frac{g}{2\pi}
\bar{\rho}_{R}\right]=0\,,
\end{eqnarray}
where the source term $j(x,t)$ is defined by Eq.(\ref{source_term}).
The solution of Eq.~(\ref{a7}) determines the phases
$\delta_\eta(t)$ according to Eqs.~(\ref{d1a}), (\ref{a7a}).
Remarkably, Eq.~(\ref{a7a}) expresses the phase $\delta_\eta(t)$
affected by the electron-electron interaction,  through the
asymptotic behavior of $\bar{\rho}(x,t)$ in the non-interacting
parts of the wire (regions I and III in Fig.~\ref{fig1}). The phases
$\delta_\eta(t)$ determine the GFs via\cite{footnote_Yuval}
\begin{equation}
\label{d3}
G^\gtrless_R(\tau)=\mp \frac{i\Lambda}{2\pi u}
\frac{\overline{\Delta}_R[\delta_R(t)]\overline{\Delta}_L[\delta_L(t)]}
{(1\pm i\Lambda \tau)^{1+\gamma}},
\end{equation}
where
\begin{equation}
\gamma=(1-K)^2/2K\,,
\end{equation}
and
\begin{equation}
K=(1+g/\pi v)^{-1/2}
\end{equation}
 is the standard LL parameter in the interacting region.

To explicitly evaluate $\delta_\eta(t)$ for the structure of
Fig.~\ref{fig1}, it is convenient to rewrite Eqs. (\ref{a7}) as a
second-order differential equation for the current
\begin{equation}
\label{current}
\bar{J}=v(\bar{\rho}_R-\bar{\rho}_L),
\end{equation}
\begin{equation}
\label{d2}
(\omega^2+\partial_x
u^2(x)\partial_x)\bar{J}(\omega,x)=0\,,\ \  x \neq 0,
\end{equation}
where
\begin{equation}
u(x)=v(1+g(x)/\pi v)^{1/2}=\frac{v}{K(x)}\,
\end{equation}
is a spatially dependent plasmon velocity.
Reflection and transmission of plasmons on both boundaries
is characterized by the coefficients $r_\eta$, $t_\eta$
($r_\eta^2+t_\eta^2 =1$); here the subscripts $\eta$ refer to the
boundaries between regions I/II and II/III. For simplicity, we assume
them to be constant over a characteristic frequency
range\cite{footnote_tau} $\omega \sim \tau^{-1}$. The scattering
matrices on the left and right boundaries have the form
\begin{equation}
\label{s-matrices}
S_L = \left(
\begin{array}{cc}
t_L & - r_L \\
r_L & t_L
\end{array}
\right), \qquad
S_R = \left(
\begin{array}{cc}
t_R &  r_R \\
- r_R & t_R
\end{array}
\right),
\end{equation}
where the first component corresponds to the left mover and the
second one to the right mover.

Solution of Eq.~(\ref{d2}) is quite straightforward. The boundary
points $x=\pm L/2$ and the observation point $x=0$ divide the $x$ axis
into four regions (I, II$_-$, II$_+$, and III). In each of the regions
the function $\bar{J}(\omega,x)$ satisfies the homogeneous wave
equation, with the velocity $v$ (in regions I and III) or $u$ (in
regions  II$_-$ and II$_+$). The solution in each of the regions is
thus a sum of two waves propagating left and right. As discussed after
Eq.~(\ref{d1}), we need an advanced solution, which imposes the
condition that in the leads (regions I and III) only incoming waves
are present. There remain six coefficients that are fixed by the
boundary conditions at the sample/lead boundaries
[see Eq.~(\ref{s-matrices})] and by the matching condition at the
observation point ($x=0$). The latter condition is 
generated by the source term in Eq.~(\ref{a7}).   

Solving Eq.~(\ref{d2}) and using Eq.~(\ref{current}),  we find
the quantum density components $\bar{\rho}_\eta$.
In accordance with Eq.~(\ref{a7a}) the scattering phases $\delta_\eta(t)$
are determined by the behavior of $\bar{\rho}_\eta$ in the asymptotic
regions ($x<-L/2$  for  $\bar{\rho}_R$ and $x>L/2$  for
$\bar{\rho}_L$). We find
\begin{eqnarray}
\label{same_points4}
\bar{\rho}_R(\omega,x)&=&\frac{(1+K)t_L}{2\sqrt{2K}v}
\frac{e^{ikx+i(k-\kappa)L/2}}{1-r_Rr_Le^{-2i\kappa L}} \left(1
-e^{i\omega \tau}\right)\nonumber \\& \times  &\left(1-
r_Rre^{-i\kappa L}\right), \qquad x < -\frac{L}{2}\,;\\
\label{same_points5}
\bar{\rho}_L(\omega,x)&=&\frac{(1+K)t_R}{2\sqrt{2K}v}
\frac{e^{-ikx+i(k-\kappa)L/2}}{1-r_Rr_L e^{-2i\kappa L}} \left(1
-e^{i\omega \tau}\right)\nonumber \\ & \times & \left(r+ r_Le^{-i\kappa
L}\right), \qquad x > \frac{L}{2}\,.
\end{eqnarray}
Here we use the notations $k=\omega/v$, $\kappa=\omega/u$, and
$r=(1-K)/(1+K)$.
Substituting this in Eq.~(\ref{a7a}), we obtain $\delta_\eta(t)$ in the
form of a superposition of rectangular pulses,
\begin{equation}
\label{delta_spinless}
\delta_\eta(t)=\sum_{n=0}^\infty \delta_{\eta,n}w_\tau(t,t_n)\,,
\end{equation}
where
\begin{equation}
t_n=(n+1/2-1/2K)L/u\,
\end{equation}
and
\begin{eqnarray}&&
\label{delta_spinless1}
\delta_{\eta,2m}=\pi t_{-\eta} r_L^mr_R^m(1+\eta K)/\sqrt{K}\,,
\nonumber \\&&
\delta_{\eta,2m+1}=
-\pi t_{-\eta} r_\eta^{m+1} r_{-\eta}^{m} (1-\eta K)/\sqrt{K}\,.
\end{eqnarray}

For the ``partial equilibrium'' state (where $n_R(t)$ and $n_L(t)$
are of Fermi-Dirac form but with different temperatures and chemical
potentials) the
functional determinants are Gaussian functions of phases,
reproducing  earlier results of functional bosonization
\cite{GGM_2009}. Indeed, using Eq.~(\ref{eq_app6}), we find
\begin{eqnarray}
\label{d3_partial_equilibrium}
G^\gtrless_R(\tau)&=&\mp \frac{i\Lambda}{2\pi u}
\frac{1}
{(1\pm i\Lambda \tau)^{1+\gamma}} \nonumber \\ &\times &
\left(\frac{\pi T_R\tau}{\sinh\pi T_R\tau}\right)^{1+\alpha}
\left(\frac{\pi T_L\tau}{\sinh\pi T_L\tau}\right)^\beta\,,
\end{eqnarray}
where the exponents $1+\alpha$ and $\beta$ are given by the sums
\begin{eqnarray}
\label{alpha_beta}
1+\alpha &\equiv& \sum_{n=0}^\infty\left(\frac{\delta_{R,n}}{2\pi}\right)^2\,,
\nonumber \\
\beta &\equiv& \sum_{n=0}^\infty\left(\frac{\delta_{L,n}}{2\pi}\right)^2 \,.
\end{eqnarray}
Substituting here the results (\ref{delta_spinless1}) for the phases
$\delta_{\eta,n}$, we obtain
\begin{eqnarray}&&
\label{alpha_beta2}
1+\alpha=
\frac{{\cal T}_L}{1-{\cal R}_L {\cal R}_R}
\bigg[1+\frac{\gamma}{2}(1+{\cal R}_R)\bigg]\,,
\nonumber \\&&
\beta=\frac{{\cal T}_R}{1-{\cal R}_L{\cal R}_R}\bigg[{\cal R}_L+\frac{\gamma}{2}
(1+{\cal R}_L)\bigg]\,,
\end{eqnarray}
in agreement with Ref.~\onlinecite{GGM_2009}.
One may check that, due to the sum rule
\begin{equation}
\label{sum_rule}
\alpha+\beta=\gamma\,,
\end{equation}
at thermal equilibrium ($T_R=T_L$)
the GFs $G^\gtrless$ are independent of plasmon
transmission/reflection amplitudes.

\begin{figure}
\includegraphics[width=0.95\columnwidth,angle=0]{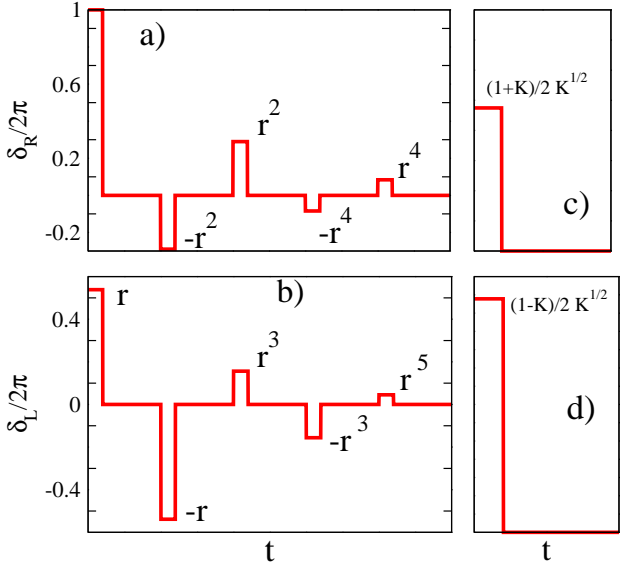}
\caption{Phases $\delta_\eta(t)$ entering Eq.~(\ref{d3})
 for the GFs for sharp [a),b); $r=(1-K)/(1+K)$] and adiabatic [c),d)]
  boundaries. }
\label{fig2}
\end{figure}

The phases $\delta_\eta(t)$ are shown in Fig.~\ref{fig2} for two
limits of adiabatic ($r_\eta=0$) and sharp,
\begin{equation}
r_\eta=(1-K)/(1+K)\,,   \nonumber
\end{equation}
boundaries. Let us stress that when we speak here about sharp
boundaries, we mean that the extension of the contact regions is small
compared to the characteristic plasmon wave length $\sim u/\omega$. It
is assumed throughout the paper that the structure is always smooth on
the scale of the electron wave length, so that no electron
backscattering takes place.

In physical terms  $\delta_\eta(t)$ characterizes phase
fluctuations in the leads that arrive  at the measurement point $x=0$
during  the time interval $[0,\tau]$.  These fluctuations govern the
dephasing and the energy distribution of electrons encoded in the
GFs $G^\gtrless_\eta(\tau)$. Up to inversion of time,
one can think of $\delta_\eta(t)$
as describing the fractionalization of a phase pulse (electron-hole
pair) injected into the wire at point $x$ during the time interval
$[0,\tau]$. This is closely related to the physics of charge
fractionalization discussed earlier
\cite{Safi,Oreg_Finkelstein,safi97,lehur,steinberg08,berg09}. At the
first step, the pulse splits into two with relative amplitudes
$(1+K)/2$ and $(1-K)/2$ carried by plasmons in opposite directions,
cf. Refs.~\onlinecite{safi97,lehur,steinberg08}.
As each of these pulses reaches the corresponding boundary, another
fractionalization process takes place: a part of the pulse is transmitted
into a lead, while the rest is reflected. The reflected pulse
reaches the other boundary, is again fractionalized there, etc.
Let us stress an
important difference between boundary fractionalization of transmitted charge
\cite{Safi,berg09} and that of dipole pulses discussed here.
While in the former case the boundaries can
always be thought of as sharp (one is dealing with the small q limit),
in the present problem the way $K(x)$
is turned on is crucially important for
reflection coefficients $r_\eta$ at $\omega\sim\tau^{-1}$.

For $\tau \ll L/u$
the coherence of plasmon scattering  may  be  neglected and
the result splits into a product
\begin{equation}
\label{c10}
\overline{\Delta}_\eta[\delta_\eta(t)]\simeq\prod_{n=0}^\infty
\overline{\Delta}_{\eta\tau}(\delta_{\eta,n})
\,,
\end{equation}
with each factor  representing a contribution of
a single phase pulse
$\delta_{\eta,n}(t)=\delta_{\eta,n}w_\tau(t,0)$.

\begin{figure}
\includegraphics[width=0.95\columnwidth,angle=0]{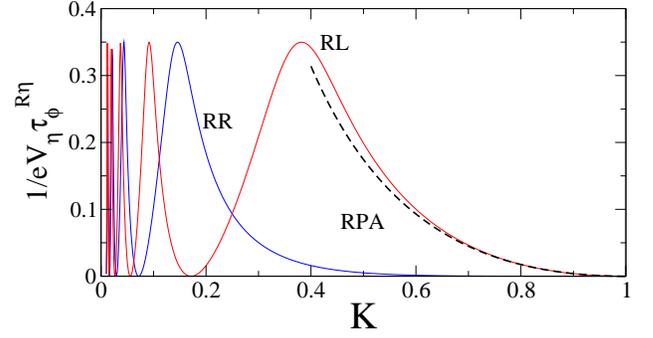}
\caption{Dephasing rates $1/\tau_\phi^{RR}$ and $1/\tau_\phi^{RL}$ as
function of LL parameter $K$ for the adiabatic case and double step
distributions with $a_R=a_L=1/3$. }
\label{fig5}
\end{figure}

We now apply our general result (\ref{d3}), (\ref{c10}) to the ``full
non-equilibrium'' case, when $n_\eta(\epsilon)$ have a double step
form, Eq.~(\ref{double-step}). To obtain the exact form of the Green
function $G_\eta(\tau)$, one has to evaluate the Toeplitz determinants
numerically. Here we restrict ourselves to the evaluation of the
long-time asymptotics of $G_\eta(\tau)$ that can be found analytically
employing Eq.~(\ref{a5}) and governs the broadening of the split
zero-bias-anomaly dips\cite{GGM2008,GGM_2009}. We focus on the
adiabatic limit when the distribution function remains unchanged and
the broadening is solely due the non-equilibrium dephasing
rate\cite{GGM2008,GGM_2009}, $1/\tau_\phi^\eta$.
We obtain
\begin{equation}
1/\tau_\phi^R=1/\tau_\phi^{RR}+1/\tau_\phi^{RL}\,,
\end{equation}
where $1/\tau_\phi^{\eta\eta'}$ is the contribution to dephasing of the $\eta$
fermions governed by the distribution of the $\eta'$ fermions. These
dephasing rates are found to be
\begin{equation}
\label{c11}
1/\tau_\phi^{R\eta}=-\frac{eV_\eta}{2\pi}\ln\left(
1-4a_\eta(1-a_\eta)\sin^2
\frac{\pi(1+\eta K}{2\sqrt{K}}\right),
\end{equation}
see Fig.~\ref{fig5}.

Two remarkable features of this result should be pointed out.
First, let us compare our results with the results of RPA
approximation. Consider the weak-interaction regime,
$\gamma\ll 1$. We then obtain
\begin{eqnarray}
&& 1/\tau_\phi^{RL}\simeq\pi\gamma eV_L a_L(1-a_L)\,, \\
&& 1/\tau_\phi^{RR}\simeq\pi(\gamma^2/8) eV_R a_R(1-a_R)\,.
\end{eqnarray}
This should
be contrasted with RPA which predicts equal $1/\tau_\phi^{RL}$ and
$1/\tau_\phi^{RR}$, see Ref.~\onlinecite{GGM2008}.
While $1/\tau_\phi^{RL}$ agrees
with the RPA result,  $1/\tau_\phi^{RR}$ is parametrically smaller
(suppressed by an extra factor of $\gamma$). The reason for this
failure of  RPA is clear from our analysis. For a weak interaction
the contributions of R and L movers to $G_R$ are given by the
functional determinants $\Delta_{\eta\tau}(\delta_\eta)$ with phases
(for adiabatic boundaries) $\delta_L\simeq (1-K)\pi$ and $\delta_R
\simeq \pi(1+K)$. While the contribution of the small phase
$\delta_L$ is captured correctly by RPA, a small-$\delta$ expansion
of $\ln\Delta_{R\tau}(\delta_R)$ fails for large $\delta_R$ (apart
from equilibrium and ``partial equilibrium'' where
$\ln\Delta_{\eta\tau}(\delta)\propto \delta^2$.)

Another important observation is that for certain values of the
interaction parameter $K$  (different for $\eta=R$ and $L$)
the dephasing rates $1/\tau_\phi^{R\eta}$ vanish.
This implies that, for these values of $K$, the GF does not decay
exponentially in time, so that the power-law ZBA anomaly is not
smeared. The absence of dephasing indicates  that for these values of
interaction the system reduces in some sense to a
non-interacting model.
As we are going to show, at these points the functional determinant can
be calculated exactly.

\subsubsection{Refermionization}
\label{sec_refermionization}

The points of no-dephasing correspond to the value of the phase
$\delta$ (argument of the functional determinant) equal to
$\delta=2\pi n$ with an integer $n$.
We will demonstrate that at these points the functional determinant
$\overline{\Delta}_\tau(\delta)$
can be calculated exactly by ``refermionization''.
The case $\delta=2\pi$ corresponds to the non-interacting ($K=1$)
single-particle GF and has been already
analyzed in Sec.~\ref{sec_free_fermions}.
To study  the case $\delta=4\pi$, we consider a two-fermion GF
\begin{equation}
\label{eq_4_fermions}
G_2=\langle T\psi^\dagger(1)\psi^\dagger(2)\psi(3)\psi(4)\rangle\,.
\end{equation}
We  focus on the limit of  merging points, $t_1=t_2=0$, $t_3=t_4=\tau$;
$x_1, x_2, x_3, x_4 \to x$,
which corresponds to simultaneous creation and annihilation
of two fermions, and thus should generate
$\overline{\Delta}_\tau(4\pi)$.
For  non-interacting electrons the GF $G_2$
can be readily calculated.
Using Wick's theorem, we find
\begin{equation}
\label{eq_four_fermions}
G_2=G(3,1)G(4,2)-G(4,1)G(3,2)\,.
\end{equation}
If  the spatial points strictly coincide,  $x_1=x_2=x_3=x_4=x$,
the function $G_2$  vanishes.
A finite result is obtained after splitting the points by
distances of the order of Fermi length, $s_i \sim v/\Lambda$.
We thus find
\begin{eqnarray}
\label{eq_four_point_merge1}
\hspace*{-1cm} G_2(\tau)& = & \frac{1}{2}(s_1-s_2)(s_4-s_3)
\nonumber \\& \times &
(\partial_{\tilde{s}_1}-\partial_{\tilde{s}_2})^2
G(\tau,\tilde{s}_1)G(\tau,\tilde{s}_2)
\mid_{\tilde{s}_1=\tilde{s}_2=0}\,,
\end{eqnarray}
where $x_i = x +s_i$.
In the bosonic description this corresponds to
\begin{eqnarray}&&
G_2(\tau) = \left(\frac{\Lambda}{2\pi v}\right)^2
\big\langle T_K e^{2i\phi(\tau)-2i\phi(0)}\big\rangle.
\end{eqnarray}
As was shown above, this correlation function can be evaluated, with
the result expressed in terms of a functional determinant,
\begin{equation}
\label{eq_four_point_merge2}
G_2(\tau) =
\left(\frac{\Lambda}{2\pi
    v}\right)^2\frac{\overline{\Delta}_\tau(4\pi)}{(\Lambda\tau)^4}\,,
\end{equation}
where we used $\tau \gg \Lambda^{-1}$.

\begin{figure}
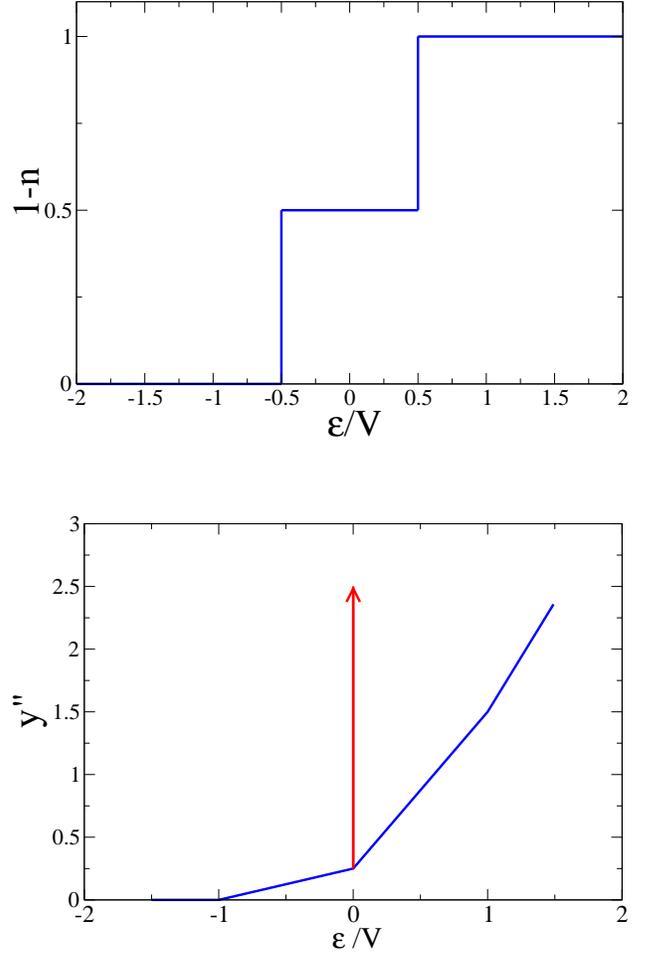

\includegraphics[width=0.95\columnwidth,angle=0]{1-n.eps}

\vspace{1cm}
\includegraphics[width=0.95\columnwidth,angle=0]{Det_4pi.eps}
\caption{Refermionization at a no-dephasing point. Upper plot:
double-step distribution function $1-n(\epsilon)$.
Lower plot: function $y''(\epsilon/V)$, where
$
G^>_{\delta_0=-\pi}(\epsilon)=-i/(2v)
(V/\Lambda)^3 y(\epsilon/V)$ is the FES Green function (\ref{b1})
proportional to the determinant ${\overline{\Delta}}_{\tau}(4\pi)$
with the argument $\delta=4\pi$ being integer multiple of $2\pi$. The
second derivative is plotted in order to emphasize singularities. The
arrow at $\epsilon=0$ denotes a delta-function contribution to $y''$.
 }
\label{distribution_at_4pi}
\end{figure}

Comparing Eqs.~(\ref{eq_four_point_merge1}) and
(\ref{eq_four_point_merge2}), we express
$\overline{\Delta}_\tau(4\pi)$ through the free-electron GFs,
\begin{equation}
\label{eq_refermionization}
\overline{\Delta}_\tau(4\pi)=
(2\pi)^2 (v\tau)^4 (\partial_{s_1}-\partial_{s_2})^2
G(s_1,\tau)G(s_2,\tau)\,.
\end{equation}
The numerical coefficient $(2\pi)^2$ was restored by comparison
with equilibrium  case.
For a double-step distribution function, Eq.~(\ref{double-step}),  we
find from Eq.~(\ref{eq_refermionization}):
\begin{eqnarray}
\label{4pi}
\!\!\!\!\!\!\!\! \overline{\Delta}_{\tau}(4\pi)&=&e^{2i(a-1)V\tau}
\left(\frac{ \pi T_\eta\tau}{\sinh\pi T_\eta\tau}\right)^2 \nonumber \\
& \times & \bigg[a_\eta(a_\eta-1)(V\tau)^2e^{iV\tau}
\nonumber \\
&+& \left(a+(1-a)e^{iV\tau}\right)^2
\left(\frac{\pi T_\eta\tau}{\sinh\pi T_\eta\tau}\right)^2\bigg].
\end{eqnarray}
We see that  $\overline{\Delta}_{\tau}(4\pi)$ shows oscillations in
$\tau$. At zero temperature there is no exponential damping.
The absence of damping is a manifestation of the vanishing dephasing rate,
see the discussion above.
Another interesting property of the result (\ref{4pi}) is
the emergence of oscillations  with three frequencies: $-2aeV$,
$(1-2a)eV$, and $(2-2a)eV$, implying three points of singular behavior
in the energy space. Let us recall that the input double-step
distribution had two such points:  $-aeV$ and $(1-a)eV$. With
increasing interaction strength the corresponding twofold singularity gets
progressively more smeared [see Eq.~(\ref{eq_80})],
but then as $\delta$ approaches $\delta=4\pi$,
a threefold singularity emerges at the new positions, see
Fig.~\ref{distribution_at_4pi}.

This  procedure can be extended to a more general case of  $\delta=2\pi n$.
Indeed,   the simultaneous creation and
annihilation of $n$ non-interacting fermions is described by
\begin{equation}
G_n(\tau)=\langle (\psi^\dagger(\tau))^n(\psi(0))^n\rangle\,,
\end{equation}
where we again imply  a point splitting
on a distance of the order of Fermi wave length, i.e. $\sim v/\Lambda$.
(One can check that the relation resulting from this consideration
does not depend on details of the point-splitting procedure).
The function $G_n$ can be expressed in terms of single-particle GFs as
follows:
\begin{eqnarray}
\label{eq_ref1}
G_n(\tau) &=&
C_ns^{n(n-1)} \prod_{i \neq j}^n
(\partial_{s_i}-\partial_{s_j})
G(s_1,\tau)
\nonumber \\
&\times &
G(s_2,\tau)\dots G(s_n,\tau)\mid_{s_i=0}\,.
\end{eqnarray}
Here $C_n$ are numerical coefficients of the order of unity.
On the other hand, in the bosonic framework we have
\begin{equation}
\label{eq_ref2}
G_n(\tau) = \left(\frac{\Lambda}{2\pi
    v}\right)^n\frac{\overline{\Delta}_\tau(2\pi
  n)}{(\Lambda\tau)^{n^2}}\,.
\end{equation}
Demanding the equivalence of Eqs.~(\ref{eq_ref1}) and (\ref{eq_ref2}),
we establish the identity
\begin{eqnarray}
\label{eq_ref3}
\overline{\Delta}_\tau(2\pi n)&=&C_n(v\tau)^{n^2}\prod_{i\neq
  j}^n(\partial_{s_i}-\partial_{s_j})\nonumber \\
&\times&  G(s_1,\tau)\dots G(s_n,\tau))\mid_{s=0}\,,
\end{eqnarray}
expressing the functional determinant
$\overline{\Delta}_\tau(\delta)$ through free fermionic GFs
$G(\tau)$ for $\delta=2\pi n$. The numerical coefficients
$C_n$ can be restored by comparison with the known result
for $\overline{\Delta}_\tau(2\pi n)$ at equilibrium.
The explicit form of  $\overline{\Delta}_\tau(\delta=2\pi n)$
can be readily found
by substituting in Eq.~(\ref{eq_ref3}) an explicit expression for the
GF for a given distribution function.

\subsubsection{Tunneling into non-interacting regions}
\label{sec_nonint_regions}

Next we discuss the tunneling spectroscopy for the non-interacting parts
of the wire. Let us focus on  the right-moving electrons;
the analysis of left-moving ones can be done in the same way.
For  $x_1,x_2 < - L/2$ (region I in Fig.~\ref{fig1}) the GF
is the one of free fermions,
as the right-moving particles emerging from the left reservoir
are not yet aware of the interacting region they are about to enter.
The situation is less trivial for  $x_1,x_2>L/2$ (region III). Indeed,
while the strength of the interaction in this region is zero,
right-moving electrons there  have passed through the interacting
part of the wire, which modifies their Green function.
We will show below that the GFs $G^\gtrless_R(x_1,t_1;x_2,t_2)$ in the
non-interacting region satisfy  Galilean invariance: they depend on
$(x_1-vt_1)-(x_2-vt_2)$ only. For this reason, it is sufficient to
consider $x_1=x_2$ to obtain the full information about the GF.

The evaluation of the GF is performed in the same way as in the
interacting region, yielding the result (\ref{d3}). The phases
$\delta_\eta(t)$ are now given by
\begin{eqnarray}
\label{eq_z1}
\delta_R(t)&=& \sum_{n=0}^\infty\delta_{R,n}w_{\tau}(t,x_1/v+2t_n)\,,\\
\label{eq_z2}
\delta_L(t)&=&2\pi r_Rw_{\tau}\left(t,\frac{x_1-L}{v}\right)\nonumber \\&+&
\sum_{n=0}^\infty\delta_{L,n}w_{\tau}\left(t,
\frac{x_1}{v}+\frac{L}{u}+2t_n\right)\,,
\end{eqnarray}
with the following amplitudes of rectangular pulses:
\begin{eqnarray}
\delta_{R,n}&=&2\pi t_L t_R (r_L r_R)^n\,,
\nonumber \\
\delta_{L,n}&=&-2\pi (r_Lr_R)^nr_Lt_R\,.
\end{eqnarray}
In the case of smooth boundaries only one pulse is created,
$\delta_{R,0}=2\pi$, reproducing the free fermion GF. Thus,
in the adiabatic case, the interaction has no influence on GFs in the
non-interacting parts of the wire, as expected.
If the transition between non-interacting and interacting parts of the wire
is not smooth, plasmon scattering takes place.
This process leads to a redistribution of electrons over energies
\cite{GGM_2009}
and thus affects GFs in the non-interacting region.

\subsection{Green functions at different points and Aharonov-Bohm  interferometry}
\label{sec_different_points}

\begin{figure}
\includegraphics[width=0.95\columnwidth,angle=0]{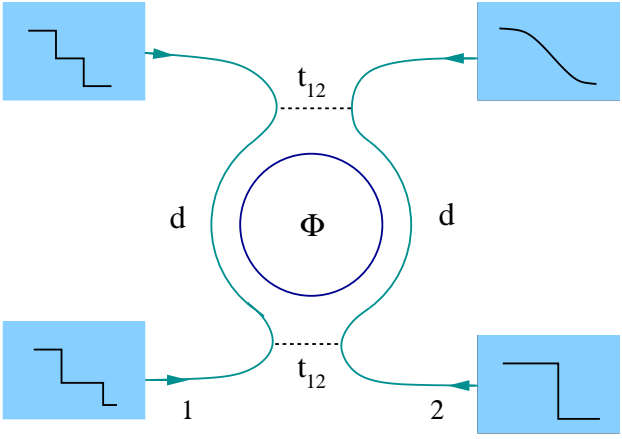}
\caption{Aharonov-Bohm setup in a four terminal geometry,
with tunnel coupling (dashed lines) at two points.
Both interferometer arms have length d.}
\label{figAB}
\end{figure}

So far we have discussed  GFs at coinciding spatial points, having
in mind tunneling spectroscopy experiments.
We now consider GFs  at different spatial coordinates.
Such  GFs are relevant to various physical quantities, in particular,
in the context  of  Aharonov-Bohm interferometry.
The similar problem in the context of chiral edge state has been
considered in Refs.~\onlinecite{Chalker,Sukhorukov}.
Let us consider a four-terminal setup formed by two quantum wires
coupled by tunneling at two points, as schematically shown in Fig.
\ref{figAB}. Each one of the quantum wires is assumed to be a LL conductor
connected to two non-interacting electrodes with arbitrary (in
general, non-equilibrium) distribution functions,
as shown in Fig.\ref{fig1}.
We are interested in the Aharonov-Bohm effect, i.e. the
dependence on the magnetic flux $\Phi$
of the electric current flowing from wire 1 into wire 2.
Consider the situation where the tunnel
coupling between the wires 1 and 2  is
weak. We also assume that both arms of the AB-interferometer have
equal length $d$ and  tunneling occurs at points located inside
the  interacting  part of the wire. The flux dependent part
of the electric current is given by
\begin{eqnarray}
\label{I_phi_1}
I_\phi &=& |t_{12}|^2\int_{-\infty}^\infty dt e^{-i\phi}
[G_2^<(d,t)G^>_1(-d,-t) \nonumber \\
&-& G_2^>(d,t)G_1^<(-d,-t)]+{\rm h.c.}\,,
\end{eqnarray}
where  the   subscripts 1 and 2  label the wire and $t_{12}$ is the
tunneling matrix element between the wires.
Separating the GF into left and right moving part, one gets
\begin{eqnarray}
\label{I_phi_2}
I_\phi&=&|t_{12}|^2\sum_\eta\int_{-\infty}^\infty dt e^{-i\phi}
[G_{2,\eta}^<(d,t)G^>_{1,\eta}(-d,-t)\nonumber \\
&-& G_{2,\eta}^>(d,t)G_{1,\eta}^<(-d,-t)]+{\rm h.c.}\,
\end{eqnarray}
where we have neglected  terms that oscillate fast with
the interferometer size $d$.

To analyze the GF $G_\eta^\gtrless(x_1,x_2,\tau)$
between two different points of a wire, we proceed in the same way as
in the case $x_1=x_2$ above.
Integration over the classical component
of the density field leads to
equations of motion for its quantum component (we choose $\eta=R$ for
definiteness),
\begin{eqnarray}&&
\label{eq_different_points1}
(-i\omega+v\partial_x)\bar{\rho}_{R,\omega}+
\partial_x\left(\frac{g}{2\pi}\bar{\rho}_{\omega}\right)=
j(\omega,x;x_1,x_2,\tau)\,,
\nonumber \\&&
(i\omega+v\partial_x)\bar{\rho}_{L,\omega}+
\partial_x\left(\frac{g}{2\pi}\bar{\rho}_\omega\right)=0\,,
\end{eqnarray}
where we have used the $(\omega,x)$ representation;
$\bar{\rho}_{\omega}=\bar{\rho}_{R,\omega}+\bar{\rho}_{L,\omega}$.
 Equations
(\ref{eq_different_points1}) differ from the earlier Eq.~(\ref{a7})
only by the source term, which now reads
\begin{equation}
\label{eq_different_points2}
j(\omega,x;x_1,x_2,\tau)=\frac{1}{\sqrt{2}}
\bigg[\delta(x-x_1)e^{i\omega \tau}-\delta(x-x_2)\bigg]\,.
\end{equation}
Solving Eq.~(\ref{eq_different_points1}), we find for $x < -L/2$
\begin{eqnarray}
\label{eq_different_points4}
\bar{\rho}_R(\omega,x)&=&\frac{(1+K)t_L}{2\sqrt{2K}v}
\frac{e^{ikx+i(k-\kappa)L/2}}{1-e^{-2i\kappa L}r_Rr_L} \nonumber \\
&\times & \bigg[e^{-i\kappa x_2}-
e^{i(\omega \tau-\kappa x_1)}
-r_Rre^{i(x_2-L)\kappa}\nonumber \\
&+&r_Rre^{i\omega \tau+i(x_1-L)\kappa}\bigg]\,.
\end{eqnarray}
Similarly, we find for $x > L/2$
\begin{eqnarray}&&
\label{eq_different_points5}
\bar{\rho}_L(\omega,x)=\frac{(1+K)t_R}{2\sqrt{2K}v}
\frac{e^{-ikx+i(k-\kappa)L/2}}{1-e^{-2i\kappa L}r_Rr_L}\nonumber
\\&& \bigg[re^{i\kappa x_2}- re^{i(\omega \tau+\kappa x_1)}\nonumber
\\&& +r_Le^{i\omega\tau-(x_1+L)\kappa}-r_Le^{-i(x_2+L)\kappa
}\bigg]\,.
\end{eqnarray}

Employing Eqs.~(\ref{a7a}), (\ref{eq_different_points4}) and
(\ref{eq_different_points5}), we obtain the following result for the
GF:
\begin{eqnarray}
\label{GF_different_points}
&& G^\gtrless_R(x_1,x_2,\tau) =
-\frac{1}{2\pi u (\pm i\Lambda)^\gamma}\nonumber \\
&& \times  \frac{\overline{\Delta}_{R}[\delta_{R}(t)]
\overline{\Delta}_{L}[\delta_{L}(t)]}{(\tau-\frac{x_1-x_2}{u}\mp
\frac{i}{\Lambda})^{1+\alpha}(\tau+\frac{x_1-x_2}{u}\mp
\frac{i}{\Lambda})^\beta}\,.
\end{eqnarray}
It is interesting to note that for spatially separated points
the scaling of GF  with time (and consequently with energy)
is affected by plasmon scattering at the boundaries between
wire and the leads.  Surprisingly, even at equilibrium the GF inside
interacting region is affected by the way interaction is turned on.
For coinciding spatial points  the  universal LL exponents,
characteristic of an infinite wire,
are restored due to the sum rule (\ref{sum_rule}).

In the long wire limit the  functional determinant splits, as before,
into a product 
\begin{equation}
\label{determinant_different_points_R}
\overline{\Delta}_R[\delta_R(t)]\simeq\prod_{n=0}^\infty
\overline{\Delta}_{R,\tau-\frac{x_1-x_2}{u}}(\delta_{R,2n})
\overline{\Delta}_{R,\tau+\frac{x_1-x_2}{u}}(\delta_{R,2n+1})
\,.
\end{equation}
Here $\delta_{\eta,n}$ are given by Eq.~(\ref{delta_spinless1}).
The calculation of $\overline{\Delta}_L$ is performed in a similar way,
yielding
\begin{equation}
\label{determinant_different_points_L}
\overline{\Delta}_L[\delta_L(t)]\simeq\prod_{n=0}^\infty
\overline{\Delta}_{L,\tau+\frac{x_1-x_2}{u}}(\delta_{L,2n})
\overline{\Delta}_{L,\tau-\frac{x_1-x_2}{u}}(\delta_{L,2n+1})
\,.
\end{equation}

We see that in the case of a GF at different spatial points the time
argument of  $\overline{\Delta}_{\eta,\tau}(\delta_\eta)$ (determining
the duration of the pulses)  is replaced as compared to the case of
$x_1=x_2$ by
\begin{equation}
\tau\rightarrow \tau \mp\eta\frac{x_1-x_2}{u}\,,
\end{equation}
with the $-$  ($+$) sign corresponding to even (respectively, odd)
pulses. It is easy to understand the reason for these alternating
signs. The even pulses are those that experience an even number
of reflections, thus preserving their chirality, while the odd pulses
experience an odd number of reflections and thus invert their
chirality.    
We note that in the case when both points are located in one of
the non-interacting regions
($x_1,x_2 >L/2$ or $x_1,x_2<-L/2$), the same consideration leads to
an analogous replacement of the time argument but with the bare
velocity $v$,
\begin{equation}
\tau\rightarrow \tau - \frac{x_1-x_2}{v}
\end{equation}
in the phases $\delta_\eta(t)$, Eqs.~(\ref{eq_z1}) and (\ref{eq_z2}),
entering  
$\overline{\Delta}_{\eta,\tau}(\delta_\eta)$  and, correspondingly, 
in GFs. Since there is no fractionalization at the tunneling processes
into a non-interacting region, only half of the pulses survives and no
sign alternation arises.

Of particular interest is the value of GF at the
interaction-renormalized ``light cone'', $x_1-x_2=\pm ut$.
The value of the GF at these points determines the integral for the
interference current in Eq.~(\ref{I_phi_2}), see also
Refs.~\onlinecite{lehur,GMP}.
Let us consider for simplicity the case of adiabatic barrier
($r_L=r_R=0$) when each of the products 
(\ref{determinant_different_points_R}),
(\ref{determinant_different_points_L})
reduces to the first factor. 
Compared to the limit of coinciding spatial points, the duration of
pulse in the  functional determinant has changed. The contribution
associated with  ($x_1-x_2= \eta ut$) leads to a doubling of
pulse duration in  $\overline{\Delta}_{-\eta}$ , while
$\overline{\Delta}_\eta$ has disappeared altogether.
As we see now, the dephasing rate governing the exponential damping
of the GF $G_\eta^\gtrless$ at $x_1-x_2=\eta' u\tau$ is
\begin{equation}
\label{tau_phi_AB}
1/\tau_{\phi (\eta')}^{AB;\eta} = 2/\tau_\phi^{\eta,-\eta'}\,,
\end{equation}
where  $1/\tau_\phi^{\eta\eta'}$ are the partial dephasing rates for
the tunneling spectroscopy problem (coinciding spatial points), as
introduced in Sec.~\ref{sec_tunnel_int}. The dephasing rates
(\ref{tau_phi_AB})
manifest themselves in the interferometry
measurements by inducing an exponential damping of the corresponding
contributions to the Aharonov-Bohm
oscillations (thus the superscript ``AB''). In the
limit of large interferometer size $d$, the contribution with the
lowest dephasing rate will dominate,
\begin{equation}
1/\tau_\phi^{AB} =
  \min_{\eta,\eta'=R,L} 2/\tau_\phi^{\eta\eta'}\,.
\label{tau_phi_AB_min}
\end{equation}

For double step distributions, the
dephasing rates (\ref{tau_phi_AB}) are given
(up to a factor of 2) by Eq.~(\ref{c11}). With
increasing interaction the dephasing rate $1/\tau_\phi^{AB}$ begins
to oscillate as a function of interaction parameter $K$, as
illustrated (for the case of adiabatic contacts with leads) in
Fig.~\ref{fig5}. This leads to  a remarkable prediction: the
visibility of Aharonov-Bohm oscillations should be a strongly
oscillating function of the interaction strength.

\subsection{Spinful Luttinger liquid}
\label{spinfull}

We now consider the problem of tunneling spectroscopy for spinful electrons.
The analysis is a straightforward extension of the spinless case,
analyzed in Sec.~\ref{spinless}.
We begin with a fermionic Hamiltonian, which, in the spinful case,
is given by
\begin{eqnarray}
H &=& H_0+H_{\rm ee}\,,
\\
H_0 &=& -iv\sum_{\sigma} \left(\psi_{R,\sigma}^\dagger\partial_x\psi_{R,\sigma}-
\psi^\dagger_{L,\sigma}\partial_x\psi_{L,\sigma}\right)\,,
\\
H_{ee}&=& \frac{1}{2}\sum_{\eta,\eta';\sigma,\sigma'}\int dx g(x)\rho_{\eta,\sigma}
\rho_{\eta',\sigma'}\,.
\end{eqnarray}
where the index $\sigma=\uparrow,\downarrow$ labels the spin projection.
We now switch to a Lagrangian description.
To construct the free part of the  action on the Keldysh contour we
repeat the steps
described in detail in Sec.~\ref{sec_free_fermions} and  find
\begin{equation}
S_0=\sum_{\eta=R,L;\sigma=\uparrow,\downarrow}\bar{\rho}_{\eta,\sigma}
\Pi^{a^{-1}}_\eta\rho_{\eta,\sigma}-i\ln
Z[\bar{\chi}_{\eta,\sigma}]\,,
\end{equation}
where
\begin{eqnarray}
\bar{\chi}_{\eta,\sigma}=\Pi_{\eta}^{a^{-1}}\bar{\rho}_{\eta,\sigma}\,.
\end{eqnarray}
The interacting part of the action reads
\begin{equation}
S_{ee}=-\sum_{\eta,\eta',\sigma,\sigma'}\int dx g(x)
\rho_{\eta,\sigma}\bar{\rho}_{\eta',\sigma'}\,.
\end{equation}

To describe the tunneling spectroscopy measurements,
we need to find the single-particle GFs
\begin{eqnarray}&&
\label{eq_spin9}
G^<_{0,\eta,\sigma}(x,t)=i\langle
\psi^\dagger_{\eta,\sigma}(0,0)
\psi_{\eta,\sigma}(x,t)
\rangle \,,
 \nonumber \\&&
G^>_{0,\eta,\sigma}(x,t)=-i\langle\psi_{\eta,\sigma}(x,t)
\psi^\dagger_{\eta,\sigma}(0,0)\rangle\,. 
\end{eqnarray}
The fermionic operators are expressed in terms of bosonic fields (which
now also carry the spin label) in the usual way,
\begin{equation}
\label{eq_spin8}
{\psi}_{\eta,\sigma}(x) \simeq \left(\frac{\Lambda}{2\pi v}\right)^{1/2}
e^{i\eta p_F x}e^{i{\phi}_{\eta,\sigma}(x)}\,.
\end{equation}
Substituting Eq.~(\ref{eq_spin8}) into Eq.~(\ref{eq_spin9}),
representing the GF as a bosonic functional integral with the action
$S_0+S_{ee}$, and performing
the integration over the classical component of the density field,  we find
the equation of motion satisfied by the quantum components of the field,
\begin{eqnarray}
\label{eq_spin2}
&& (\partial_t+v\partial_x)\bar{\rho}_{R}+
\partial_x\frac{g}{2\pi}
(\bar{\rho}_{R}+\bar{\rho}_{L})
=j(x,t)\,,
\nonumber \\&&
(-\partial_t+v\partial_x)\bar{\rho}_{L}+\partial_x\frac{g}{2\pi}
\left(\bar{\rho}_{R}+\bar{\rho}_{L}\right)=0\,,
\end{eqnarray}
and
\begin{eqnarray}&&
\label{eq_spin3}
(\partial_t+v\partial_x)\bar{s}_{R}=j(x,t)\,,
\nonumber \\&&
(-\partial_t+v\partial_x)\bar{s}_{L}=0\,.
\end{eqnarray}
Here we have passed to new variables that describe the
spin and charge sectors of excitations,
\begin{eqnarray}&&
\label{eq_spin1}
\bar{\rho}_R=\bar{\rho}_{R,\uparrow}+\bar{\rho}_{R,\downarrow}\,, \ \
\bar{\rho}_L=\bar{\rho}_{L,\uparrow}+\bar{\rho}_{L,\downarrow} \nonumber \\&&
\bar{s}_R=\bar{\rho}_{R,\uparrow}-\bar{\rho}_{R,\downarrow}\,, \ \
\bar{s}_L=\bar{\rho}_{L,\uparrow}-\bar{\rho}_{L,\downarrow}\,.
\end{eqnarray}
As one sees, the equations for the charge and spin degrees of freedom
are decoupled, which is a manifestation of spin-charge separation.
The spin-density component  obeys the same equation as the density of
free fermions, Eq.~(\ref{d1}). Therefore, the spin sector is
characterized by a LL parameter $K_s=1$.
As follows from Eq.(\ref{eq_spin3}), the
spin component $s_\eta$ propagates through the wire without any
reflection.

To find the charge component, we define the charge density current
\begin{equation}
\label{eq_spin5}
\bar{J}=v(\bar{\rho}_R-\bar{\rho}_L)\,.
\end{equation}
In terms of $\bar{J}$,  equations (\ref{eq_spin2}) are reduced to a
second order differential equation,
\begin{equation}
\label{eq_spin6}
(\omega^2+\partial_x u^2_c\partial_x)\bar{J}=0 \qquad {\rm for}\,\, x\neq 0,
\end{equation}
 where
\begin{eqnarray}
\label{eq_spin4}
u_c^2(x)&=& v^2/K^2_c(x)\,, \nonumber \\
K_c&=&\left(1+\frac{2g}{\pi v}\right)^{-1/2}.
\end{eqnarray}
Equation (\ref{eq_spin6}) coincides with Eq.~(\ref{d2}) up to a
different definition of the LL parameter.
The interaction parameter $\gamma=(1-K_c)^2/2K_c$ and the
transmission and reflection amplitudes are determined as
for spinless fermions, with the replacement $K\rightarrow K_c$.

The resulting expression for the Green function of spinful fermions
within the non-equilibrium bosonization approach reads
\begin{equation}
\label{eq_spin7}
G^\gtrless_{R,\uparrow}(\tau)=\mp\frac{i\Lambda}{2\pi \sqrt{uv}}
\frac{\prod_{\eta,\sigma}
\overline{\Delta}_{\eta,\sigma}[\delta_{\eta,\sigma}(t)]
}
{(1\pm i\Lambda \tau)^{1+\gamma/2}}.
\end{equation}
Here we have assumed for generality that distribution functions of
spin-up and spin-down particles may be different. Therefore, the
distribution function is labeled by two indices (chirality and spin
projection); these indices are inherited by the functional determinant.
The time-dependent phases of the spinful fermions $\delta_{\eta,\sigma}(t)$
are expressed in terms of the scattering phases $\delta_\eta(t)$
of spinless fermions, Eq.~(\ref{delta_spinless}),
in the following way:
\begin{eqnarray}&&
\delta_{L,\uparrow}(t)=\delta_{L,\downarrow}(t)=
\frac{1}{2}\delta_L(t)\, ,
\label{delta_L_spin} \\&&
\delta_{R,\uparrow}(t)=\frac{1}{2}\left(\delta_R(t)+\delta^0_R(t)\right)\,,
\label{delta_R_spin_up} \\&&
\delta_{R,\downarrow}(t)=\frac{1}{2}\left(\delta_R(t)-\delta^0_R(t)\right)\,,
\label{delta_R_spin_down}
\end{eqnarray}
where $\delta^0_R(t)$ corresponds to non-interacting fermions and
consists of a single pulse with an amplitude $2\pi$,
$\delta^0_R(t)= 2\pi w_\tau(t,0)$.

We conclude that the inclusion of spin changes the scattering phases
in an essential way. This is most importantly seen when considering
the first pulse propagating to the right. Let us assume that
there is no reflection at the boundaries with non-interacting leads.
The corresponding scattering phases are each a superposition of the
spin and charge modes, see
Eqs.~(\ref{delta_R_spin_up}), (\ref{delta_R_spin_down}). Since the
velocities of these modes are different ($v$ and $u$, respectively),
then for sufficiently long wires, $L(v^{-1}-u^{-1})/\tau \gg 1$, the
first pulse splits into a charge and a spin parts. For a short wire,
the  spin pulse and the first charge pulse overlap. In this case one
has to deal with the general formula (\ref{eq_spin7}) and with
time-dependent phase containing both, spin and,
charge contributions. Hence, if the wire is sufficiently short (or,
in other words, for a given length of the wire the interaction is
sufficiently weak) the spin-charge separation does not have enough
time to develop. For
sufficiently long wires, spin-charge separation does take place, in
which case the respective determinants can be written as products of
spin and charge contributions. This decomposition is not valid for
short wires (or, for a given length of the wire, for sufficiently
weak interaction). Note that at equilibrium there is significant
simplification. The GFs depend only on the sum of the scattering
phases squared, see Eq.~(\ref{d3_partial_equilibrium}).
Due to the sum rule (\ref{sum_rule}) this combination remains
unchanged, regardless of whether spin and charge pulses overlap or
not. Thus, at equilibrium one can always think about these two modes
separately and additively. Out of equilibrium, the dependence of the
GF on scattering phases is more subtle, and the  results for
overlapping and separated pulses are different. Therefore, the
spin-charge separation occurs in this case only  for sufficiently long
wires. Focusing on this regime, we find
\begin{eqnarray}
\label{det_spin_R_up}
\overline{\Delta}_{R,\uparrow}[\delta_{R,\uparrow}] &=&
\overline{\Delta}_{R,\tau,\uparrow}(\pi) \prod_{n=0}^\infty
\overline{\Delta}_{R,\tau,\uparrow}\bigg(\frac{\delta_{R,n}}{2}\bigg),
\\
\overline{\Delta}_{R,\downarrow}[\delta_{R,\downarrow}] &=&
\overline{\Delta}_{R,\tau,\downarrow}(-\pi)
\prod_{n=0}^\infty
\overline{\Delta}_{R,\tau,\downarrow}\bigg(\frac{\delta_{R,n}}{2}\bigg),
\label{det_spin_R_down} \\
\overline{\Delta}_{L,\sigma}[\delta_{L,\sigma}]&=&
\prod_{n=0}^\infty
\overline{\Delta}_{L,\tau,\sigma}
\bigg(\frac{\delta_{L,n}}{2}\bigg).
\label{det_spin_L}
\end{eqnarray}
The first factor in each of Eqs.~(\ref{det_spin_R_up}) and
(\ref{det_spin_R_down}) originate from the spin mode yielding the phase
$\pi$, i.e. a half of the free-fermion phase value.
The scattering phases of other pulses (originating from the charge
mode) have a half of their values for spinless electrons.

\begin{figure}
\includegraphics[width=0.95\columnwidth,angle=0]{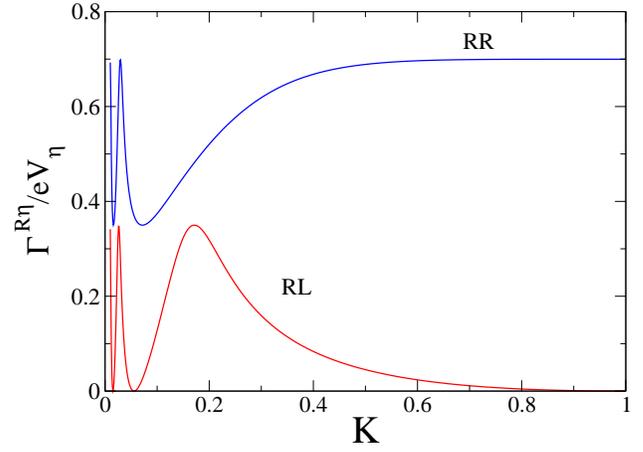}
\caption{Decay rates $\Gamma^{RR}$ and $\Gamma^{RL}$ (governing
  smearing of tunneling spectroscopy singularities) as
functions of LL parameter $K_c$ for spinful fermions.
The adiabatic coupling to leads and the double-step
distributions with $a_R=a_L=1/3$ are assumed. }
\label{fig7}
\vspace*{-0.7cm}
\end{figure}

Let us analyze this result.
Consider the case of smooth (adiabatic) contacts with leads, so that
only one pulse passes in each direction (all $\delta_{\eta,n}$ with
$n\ge 1$ are zero).
For the case of partial equilibrium, the determinants can be evaluated
explicitly, yielding
\begin{eqnarray}&&
\prod_{\sigma=\uparrow,\downarrow}
\overline{\Delta}_{R,\sigma}[\delta_{R,\sigma}(t)]=
\left(\frac{\pi T_R\tau}{\sinh\pi T_R\tau}\right)^{1+\alpha/2}\,,
\nonumber \\&&
\prod_{\sigma=\uparrow,\downarrow}
\overline{\Delta}_{L,\sigma}[\delta_{L,\sigma}(t)]=
\left(\frac{\pi T_L\tau}{\sinh \pi T_L\tau}\right)^{\beta/2}\,,
\end{eqnarray}
where $\alpha,\beta$ are given by Eq.~(\ref{alpha_beta2}).
For a double-step distribution function,
a semiclassical limit of the determinants (\ref{det_spin_R_up}),
(\ref{det_spin_R_down}),
(\ref{det_spin_L})  can be readily  evaluated.
In the large-time limit the behavior of the r.h.s. of
Eqs.~(\ref{det_spin_R_up}), (\ref{det_spin_R_down}),
(\ref{det_spin_L}) is exponential,
yielding the partial decay rates
\begin{eqnarray}
\Gamma^{RR}&=&-\frac{eV}{\pi}\bigg[\ln\left(1-4a(1-a)\right)\nonumber \\
&+&\sum_{n=0}^\infty\ln\left(1-4a(1-a)
\sin^2\frac{\delta_{R,n}}{4}\right)\bigg]\,, \nonumber\\
\Gamma^{RL}&=&-\frac{eV}{\pi}
\sum_{n=0}^\infty\ln\left(1-4a(1-a)\sin^2\frac{\delta_{L,n}}{4}\right)\bigg]\,,
\nonumber \\
&&
\label{Gamma-spinful}
\end{eqnarray}
and the total rate $\Gamma^R = \Gamma^{RR} + \Gamma^{RL}$.
Let us stress an important difference with the spinless case. There,
for smooth boundaries, the distribution function was not affected, and
the decay rate (inducing the smearing of singularities in tunneling
spectroscopy) was solely due to dephasing. In the spinful case the
situation is different: independently of the shape of the boundary the
spin-charge separation affects the distribution function of
electrons. Indeed, imagine that we perform the tunneling spectroscopy of
right-movers in the right lead (non-interacting region III of Fig.~\ref{fig1})
for the case of adiabatic boundaries. Then the phases are
$\delta_{L,n}=\delta_{R,n\ne 0}=0$ and $\delta_{R,0}=2\pi$. In the
spinless case this implied that the distribution function remained
unchanged. This is not so in the spinful situation, however: according
to Eqs.~(\ref{det_spin_R_up}), (\ref{det_spin_R_down}) we get now a
product of four determinants with arguments $\pm\pi$,
\begin{equation}
\label{QHE-det}
[\overline{\Delta}_{R,\tau,\uparrow}(\pi)]^2
\overline{\Delta}_{R,\tau,\downarrow}(-\pi)
\overline{\Delta}_{R,\tau,\downarrow}(\pi)\,,
\end{equation}
implying that the distribution function has changed. This effect
remains finite even in the limit of vanishing interaction ($K\to 1$)
as long as the long-wire  condition, $LT(u^{-1}-v^{-1}) \gg 1$, is
satisfied. Returning to the spectroscopy of the interacting region, we
conclude that both effects---dephasing and change of the
distribution function---are necessarily present in the spinful LL
case and cannot be easily ``disentangled''. The
decay rates $\Gamma$
presented in Eq.~(\ref{Gamma-spinful}) and in Fig.~\ref{fig7} yield
the combined effect of interaction on the GF $G^\gtrless_R(\tau)$ and
determine the smearing of tunneling spectroscopy singularities in the
energy space.

Finally we discuss the extension of Eq.~(\ref{eq_spin7}) to the case of GFs at
different points. In this case we find
\begin{eqnarray}&&
G^\gtrless_{R,\uparrow}(x_1,x_2,\tau)=\mp\frac{i\Lambda}{2\pi\sqrt{uv}}
\nonumber\\
&&\times \overline{\Delta}_{R,\uparrow,\tau-\frac{x}{v}}(\pi)\nonumber
\overline{\Delta}_{R,\downarrow,\tau-\frac{x}{v}}(-\pi)\\&&
\times \frac{
\prod_{n=0}^\infty\overline{\Delta}_{R,\uparrow,\tau_n^R}
\left(\frac{\delta_{R,n}}{2}\right)
\prod_{n=0}^\infty\overline{\Delta}_{R,\downarrow,\tau_n^R}
\left(\frac{\delta_{R,n}}{2}\right)
}
{\left(1\pm i\Lambda\left(\tau-\frac{x}{v}\right)\right)^{1/2}
\left(1\pm i\Lambda\left(\tau-\frac{x}{u}\right)\right)^{1/2}} \nonumber \\&&
\times\frac{
\prod_{n=0}^\infty\overline{\Delta}_{L,\uparrow,\tau_n^L}
\left(\frac{\delta_{L,n}}{2}\right)
\prod_{n=0}^\infty\overline{\Delta}_{L,\downarrow,\tau_n^L}
\left(\frac{\delta_{L,n}}{2}\right)
}{\left(1\pm i\Lambda\left(\tau-\frac{x}{u}\right)\right)^{\alpha/2}
\left(1\pm i\Lambda\left(\tau+\frac{x}{u}\right)\right)^{\beta/2}
}\,,\nonumber \\
&& \label{GF_spinfull_xt}
\end{eqnarray}
where $x=x_1-x_2$ and $\tau_n^\eta=\tau+\eta(-1)^{n+1}x/u$.
As for the spinless case, Eq.~(\ref{GF_different_points}),
the scaling of GFs with
spatially separated points is affected by plasmon scattering at the
boundaries between the interacting regions and the leads even at
equilibrium.

Before concluding this section, we point out a connection between the
spinful LL and the problem of the integer quantum Hall edge with two
edge channels (corresponding to two Landau levels below the Fermi
energy in the bulk). Non-equilibrium properties and quantum coherence
of such a system are currently attracting large interest, in
particular, in connection with experiments on quantum Hall
Mach-Zehnder interferometers \cite{MachZehnder}. In the quantum Hall
setup the edge channel index plays a role of spin. The main difference
is that, under conventional circumstances (if one does not make
special efforts to couple
counterpropagating edge modes), the quantum Hall system is chiral:
there are, say, only right-moving modes and no left-movers. This leads
to a number of essential simplifications: (i) the tunneling density
of states becomes trivial (no ZBA); (ii) the charge
fractionalization is absent (since plasmons can move only in one
direction). What remains is the charge-spin separation. This implies
that following simplifications with functional determinants
(\ref{det_spin_R_up}), (\ref{det_spin_R_down}),
(\ref{det_spin_L}) the product of which determined the tunneling
spectroscopy
Green function $G^\gtrless_{R,\uparrow}(\tau)$ within our analysis.
First, the left-mover determinants (\ref{det_spin_L})
are now absent. Second, out of the set of phases $\delta_{R,n}$ only
the $n=0$ phase remains, being equal to its free-fermion value,
$\delta_{R,0}=2\pi$. Therefore, the product of determinants takes the form
(\ref{QHE-det}) (that has appeared above in the context of a GF in the
non-interacting region of a non-chiral spinful wire with adiabatic
contacts).
The analogous statement holds for the Green function with different
spatial points, Eq.~(\ref{GF_spinfull_xt}). Specifically, for the case of the
two-channel chiral setup (relevant in the quantum Hall Mach-Zehnder
interferometry context \cite{MachZehnder}) the  last fraction (having
L determinants in the numerator) in Eq.~(\ref{GF_spinfull_xt})
disappears, while in the preceding-to-it fraction one should keep
only $n=0$ factor and set $\delta_{R,0} = 2\pi$.
An equivalent result was obtained by a
different method in the recent work \cite{Sukhorukov}.

\section{Summary}
\label{Summary}

Let us summarize the main results of this work, following the flow  of
our presentation in the paper.

\begin{enumerate}

\item
We have developed a nonequilibrium bosonization approach and
derived a bosonic theory describing the LL of
interacting 1D electrons  out of equilibrium. The theory is
characterized by an action depending on density fields defined on the
Keldysh time contour. In contrast to the
equilibrium case, this theory is not Gaussian, which is a
manifestation of the fact that the density matrix is non-diagonal in
the bosonic Fock space. We have used this theory to calculate the
electronic GFs governing various observables.

\item
We have first calculated the GF
of non-interacting fermions from our
non-equlibrium bosonization approach. The GF  is expressed in terms of
a functional determinant of the Fredholm (more specifically, Toeplitz)
type similar to those that have earlier appeared in the context of
counting statistics. The key difference is that in the case of counting
statistics the determinant is non-analytic and $2\pi$-periodic in the
counting field
(which reflects charge quantization), while in our theory the
determinant should be understood as an analytically continued function. We have
found that the free-fermion GF is described by the determinant exactly
at the point $2\pi$, which is related to the fact that in the bosonic
theory a fermion is represented by a $2\pi$-soliton.

\item
We have next generalized the GF calculation to the problem of
non-equilibrium Fermi
edge singularity describing excitation of an electron into the
conduction band within the process of photon absorption, accompanied by
creation of a core hole. The result is obtained in terms of the same
functional determinant as in the free-fermion case but the argument is
now shifted from $2\pi$ by twice the scattering phase on the core
hole.

\item
We have then applied our formalism to the problem of
interacting 1D fermions. We have considered a model of a LL wire
coupled to non-interacting 1D leads, with the interaction strength
``turned on'' in specified fashion at the boundary between the wire and
each of the leads. We have shown that the electron GFs---which
describe tunneling spectroscopy measurements---are again expressed in
terms of Fredholm determinants. The phases $\delta_\eta(t)$ entering
the expressions for the corresponding operators have a physical
interpretation in terms of fractionalization processes taking place
during the tunneling event, near the boundaries. If the
characteristic energy scales for the tunneling spectroscopy are large
compared to the inverse flight time through the LL wire (Thouless
energy)---which means that we are considering the truly 1D (rather
than 0D) regime---the functions  $\delta_\eta(t)$ represent a sequence
of rectangular pulses separated by large intervals. As a result, the
Fredholm determinant splits into a product of Toeplitz determinants of
the same type as in the cases of non-interacting fermions and the
Fermi edge singularity.

\item
We have analyzed the long-time asymptotics of the determinant
which yields the dephasing rate controlling the smearing of LL
tunneling singularities (zero-bias anomaly).
The dephasing rate for the GF of electrons
with $\eta$ ($\pm 1$) chirality is a sum of two terms
$\sum_{\eta'=\pm 1}1/\tau_\phi^{\eta\eta'}$ originating from
functional determinants which depend  on the distribution function of
left- ($\eta'=-1$) and right- ($\eta'=1$) moving electrons,
respectively. For the case of double-step distributions,
there are two important findings:

(i) At weak interaction, comparing our exact results with
those of the RPA, we find that while $1/\tau_\phi^{\eta,-\eta}$ is
correctly obtained (to  leading order) within  RPA, the RPA
result for $1/\tau_\phi^{\eta,\eta}$ is parametrically wrong. This
demonstrates that even for a weak interaction a naive perturbative
expansion (leading to RPA) may be parametrically incorrect in
LL out of equilibrium.

(ii) Both  $1/\tau_\phi^{\eta,\pm\eta}$ are oscillatory functions of
the interaction strength (or, equivalently, LL parameter $K$).
Furthermore, each of them vanishes at
certain values of $K$. At these
values the ``counting phase'' for the
corresponding determinant becomes an integer multiple of $2\pi$.
We have calculated the determinants at these no-dephasing points by a
refermionization procedure.

\item We have generalized the above results to the case of a GF with
two different spatial arguments. When considering  the value of
the GF $G_\eta$ at its main peak, $x_1-x_2=\eta ut$, the dephasing rate
is $2/\tau_\phi^{\eta,-\eta}$, while  $1/\tau_\phi^{\eta,\eta}$ does
not contribute (and thus RPA is restored for weak interaction).
The situation is reversed for the value of the $G_\eta$ at the other
peak, $x_1-x_2=-\eta ut$, where  the dephasing rate
is $2/\tau_\phi^{\eta,\eta}$. Such GFs (with $x_1-x_2=\pm \eta ut$)
enter the expression for the
interference contribution to current in an
Aharonov-Bohm interferometer formed by two  LLs coupled by tunneling
at two points. Our results imply that the dephasing rate in such a
non-equilibrium LL interferometer (and thus the visibility of
Aharonov-Bohm oscillations) is an oscillatory function of the
interaction strength.

\item
We have also considered the case of a spinful LL. The general
structure of the results for the GFs is similar, with the key difference
being that now we encounter products of determinants with phase
arguments corresponding
to the spin and charge sectors. This is a manifestation of the
spin-charge separation. One important consequence is that
the temporal decay rate of the Green function (and thus the smearing
of singularities in the tunneling spectroscopy) remains finite in the limit
of vanishing interaction strength, assuming the limit of the large system
size is taken first.  With increasing interaction strength, the
dephasing rate oscillates, similarly to the case of spinless fermions.

\end{enumerate}

The non-equilibrium bosonization formalism developed in this work has
a variety of further applications. They include, in particular,
counting statistics of charge transfer in an interacting 1D system
away from equilibrium, analysis of many-body entanglement, quantum wires with
several channels, etc. Generalizations or modifications of our
formalism should be useful for a number of  further prospective
research directions, such as systems of cold atoms and fractional
quantum Hall edges away from equilibrium.

\section{ Acknowledgments}

We thank D.~Abanin, D.~Bagrets, L.~Glazman, D.~Ivanov,
L.~Levitov, Y.~Nazarov, P.~Ostrovsky, E.~Sukhorukov,
and P.~Wiegmann for useful discussions.
Financial support by German-Israeli Foundation, Einstein Minerva
Center, US-Israel Binational Science Foundation, Israel Science Foundation,
Minerva Foundation, SPP 1285 of the
Deutsche Forschungsgemeinschaft, EU project GEOMDISS, and
Rosnauka grant 02.740.11.5072 is gratefully acknowledged.

\appendix
\section{Equilibrium:  Green functions $G_0^\gtrless$,  $G^\gtrless$
  via bosonization and Fredholm determinants  $\Delta_\tau$ }
\label{appendix1}
GFs of free fermions at equilibrium can be readily found.
Since at equilibrium the bosonic  action is Gaussian,
the functional integration over bosonic fields is straightforward.
The fermionic GFs is thus expressed as
\begin{eqnarray}
\label{eq_app1}
G^\gtrless_0(\tau)=\mp\frac{i\Lambda}{2\pi v} e^{J^\gtrless(\tau)}\,
\end{eqnarray}
in terms of the bosonic correlation functions
\begin{eqnarray}
\label{eq_app2}&&
J^>(\tau)=
-\frac{1}{2}
\langle T_K[{\phi}_{+,\eta}(0,0)-{\phi}_{-,\eta}(0,\tau)]^2\rangle\,,
\nonumber \\&&
J^<(\tau)=
-\frac{1}{2}
\langle T_K[{\phi}_{-,\eta}(0,0)-{\phi}_{+,\eta}(0,\tau)]^2\rangle\,.
\end{eqnarray}
Explicitly calculating the correlations functions of the bosonic fields
one finds
\begin{eqnarray}
\label{eq_app3}
J^\gtrless(\tau)&=&-\int_0^\infty\frac{d\omega}{\omega}e^{-\omega/\Lambda}
\nonumber \\
&\times&
\bigg[
(1-\cos\omega\tau)\coth\frac{\omega}{2T}\pm i\sin\omega\tau\bigg]\,.
\end{eqnarray}
Next we calculate  the integrals that appear on the r.h.s. of
Eq.(\ref{eq_app3}). The integral with $\sin\omega\tau$ yields
\begin{eqnarray}&&
\label{eq_app5}
\int_0^\infty\frac{d\omega}{\omega}e^{-\omega/\Lambda}
\sin\omega\tau=\arctan\Lambda\tau\,.
\end{eqnarray}
The remaining integral can be split into two parts:
\begin{eqnarray}
\label{eq_app5b}
\int_0^\infty \frac{\omega}{\omega}e^{-\omega/\Lambda}(1-\cos\omega
\tau)=\frac{1}{2}\ln(1+\Lambda^2\tau^2)\,
\end{eqnarray}
and
\begin{eqnarray}
\label{eq_app5a}
\int_0^\infty \frac{d\omega}{\omega}e^{-\omega/\Lambda}
\bigg(\coth\frac{\omega}{2T}-1\bigg)(1-\cos\omega \tau) &&
 \nonumber \\
\simeq \ln\frac{\sinh \pi T\tau}{\pi T\tau}\,;&&
\end{eqnarray}
in the second one we have dropped the convergence factor, $e^{-\omega/\Lambda}$,
which is justified in view of $T \ll \Lambda$.
Employing Eqs.~(\ref{eq_app5}), (\ref{eq_app5b}), and
(\ref{eq_app5a}), one gets
\begin{eqnarray}
\label{eq_app4}
J^\gtrless(\tau)=\ln\left(\frac{\pi T\tau}
{\sinh \pi T\tau}\frac{1}{1\pm i\tau\Lambda}\right)\,.
\end{eqnarray}
Substituting Eq.~(\ref{eq_app4}) into Eq.~(\ref{eq_app1}),
one recovers the result for the GF of free fermions,
Eq.~(\ref{Green_function_equilibrium}).

We proceed with GFs for FES at equilibrium,
\begin{equation}
G^\gtrless(\tau)=\mp\frac{i\Lambda}{2\pi
  v}\exp\left\{(1-\delta_0/\pi)^2J^\gtrless(\tau)\right\}\,.
\end{equation}
Next we relate the GF and the functional determinant $\Delta_\tau(\delta)$.
At equilibrium the latter can be evaluated as follows:
\begin{eqnarray}
\ln\Delta_\tau(\delta) &=&
-\left(\frac{\delta}{2\pi}\right)^2
\int_0^\infty\frac{d\omega}{\omega}e^{-\omega/\Lambda}
\nonumber \\
&\times&  (1-\cos\omega\tau)\coth\frac{\omega}{2T}\,.
\end{eqnarray}
Using Eq.~(\ref{eq_app5}), we find
\begin{equation}
\label{eq_app6}
\Delta_\tau(\delta)=\left(\frac{\pi\tau T}{\sinh \pi \tau T}\right)^
{(\delta/2\pi)^2}
\frac{1}{(1+\tau^2\Lambda^2)^{\frac{1}{2}(\delta/2\pi)^2}}\,.
\end{equation}
Comparing  Eq.~(\ref{eq_app6}) with Eq.~(\ref{eq_app4}),  we establish
the exact relation (including the proportionality factor) between the
GF and the functional determinant,
\begin{equation}
G^\gtrless(\tau)=\mp\frac{i\Lambda}{2\pi v}\left(\frac{1\mp
    i\Lambda\tau}{1\pm
    i\Lambda\tau}\right)^{\frac{1}{2}(1-\delta_0/\pi)^2}
\Delta_\tau(2\pi-2\delta_0)\,.
\end{equation}
We notice that the determinant
$\Delta_\tau(\delta)$ as given by Eq.~(\ref{eq_app6})
is a product
of the temperature dependent and independent parts.
It is
convenient to normalize the result by its zero temperature value,
\begin{equation}
\Delta_{\tau,T=0}(\delta)=
\frac{1}{(1+\tau^2\Lambda^2)^{\frac{1}{2}(\delta/2\pi)^2}}\,.
\end{equation}
We thus present  Eq.~(\ref{eq_app6}) in the form
\begin{equation}
\Delta_\tau(\delta)=\frac{{\overline\Delta}_\tau(\delta)
}{(1+\tau^2\Lambda^2)^{\frac{1}{2}(\delta/2\pi)^2}}\,,
\end{equation}
where ${\overline\Delta}_\tau(\delta)$ is the normalized determinant,
\begin{equation}
{\overline \Delta}_\tau(\delta)=
\left(\frac{\pi \tau T}{\sinh \pi \tau T}\right)^
{(\delta/2\pi)^2}\,.
\end{equation}
By construction, ${\overline \Delta}_\tau(\delta)=1$ for $T=0$.
It turns out to be more convenient to deal with the normalized
determinant, since all ultraviolet divergences
($\Lambda$-dependent factor) are excluded from this quantity.

\section{High-order vertices for 1D fermions: Diagrammatics}
\label{appendix2}
In this Appendix we briefly sketch an explicit calculation of  third-order
fermionic vertices by means of diagramatic fermionic approach.
Consider  the third-order vertex shown in Fig.\ref{DL}b,
\begin{eqnarray}
{\cal S}_{3,\eta}(1,2,3) &=& \langle
T_K\rho_\eta(1)\rho_\eta(2)\rho_\eta(3)\rangle
\nonumber \\
&=& \langle
\psi^\dagger_{1,\eta}\psi_{1,\eta}\psi^\dagger_{2,\eta}
\psi_{2,\eta}\psi^\dagger_{3,\eta}\psi_{3,\eta}\rangle\,.
\end{eqnarray}
Here the index $i=1,2,3$ includes the corresponding spatial coordinate ($x_i$),
time ($t_i$) and  Keldysh index $s_i$ that labels upper and lower
branches.
Using Wick theorem, we find
\begin{eqnarray}
-i{\cal S}_\eta(1,2,3)&=&G_\eta(1,3)G_\eta(3,2)G_\eta(2,1)\nonumber \\
&+&G_\eta(1,2)G_\eta(2,3)G_\eta(3,1)\,.\nonumber
\end{eqnarray}
We choose first the following  combination of Keldysh indices:
$s_1=+,s_2=+,s_3=-$.
Using Eq.~(\ref{definitions}) and passing into energy-momentum
representation,  we find
\begin{eqnarray}
\label{S3}
&& {\cal S}_{3,\eta}(\omega_1,q_1,+;\omega_2,q_2,+;\omega_3,q_3,-)
\nonumber\\
&&\quad =
-\frac{i}{v}(2\pi)^2\delta(A_1)\delta(A_2)\nonumber \\&&
\quad\times\int \frac{d\epsilon}{2\pi}n_\eta(\epsilon)
[1-n_\eta(\epsilon+\omega_1+\omega_2)]\nonumber \\&&
\quad\times[1-n_\eta(\epsilon+\omega_1)-
n_\eta(\epsilon+\omega_2)]\,.
\end{eqnarray}
Here $A_i=\omega_i-\eta vq_i$ and
$\omega_3=-\omega_1-\omega_2$,
$q_3=-q_1-q_2$.
Therefore, the third-order correlation function is restricted to the
light-cone with respect to all its coordinates. At equilibrium the
integration over energy yields zero, making  the correlation function
vanish. On the other hand, in the non-equilibrium situation the result
is in general non-zero.
Repeating the calculations for all possible  choices of Keldysh
indices $s_1,s_2,s_3$, we find that all third-order vertices are
equal and given by Eq.~(\ref{S3}). When transformed from $s=\pm$ to
quantum and classical components, this means that the result is
non-zero only when all indices are classical.
Thus, the explicit  diagrammatic calculations of the third-order vertex
confirms that (i) it is restricted to the light-cone, and (ii) only
correlations of classical fields are non-zero. We have checked this
also for vertices of fourth order. These results are in full agreement
with the general treatment (valid for vertices of all orders)
performed in Sec.~\ref{sec_free_fermions}.


\begin{thebibliography}{99}
\bibitem{Klein_factors} We omit Klein factors, as they are of no
  importance for our problem.
\bibitem{note-DL} Equation (11) of
  Ref.~\onlinecite{DzLar:73} is different from our Eq.~(\ref{DL-correlators})
  in that the former contains a product of kinematic factors
  $\prod_{i=1}^n(\omega_i-\eta vq_i)$ corresponding to all $n$
  vertices. This would allow contributions that are of mass-shell type
  with respect to just one of the vertices. However, one can in fact
  derive a much stronger equation, Eq.~(\ref{DL-correlators}), that
  imposes the mass-shell restriction with respect to all vertices
  simultaneously.
\bibitem{anomalous}
Due to the linearization of the electron spectrum
special care needs to be taken in order to account for the anomalous
contribution to
the partition function $Z$ (related to the Schwinger anomaly), as
has been also mentioned in Sec.~\ref{sec_DL}. This contribution is
given by the first term in the exponent on the r.h.s. of
Eq.~(\ref{identity}). This term does not depend on the fermionic distribution
and can be straightforwardly obtained in
the standard diagrammatic analysis performed at equilibrium.
We restore it in the final result for $Z_\eta(V,\bar{V})$,
Eq.~(\ref{eq_det2}).
\bibitem{footnote_Yuval}
The subscript $\eta$ in the notation for phases $\delta_\eta(t)$ corresponds
to the chirality of the corresponding component $\bar{\rho}_\eta$ of the
saddle-point solution and to the chiral index of the Fredholm determinant
$\bar{\Delta}_\eta$.   Clearly, the saddle-point solution (and thus
$\delta_\eta$) depends also on chirality of the Green function calculated.
For definiteness we focus on $G_R$ and drop the additional index (which will
be "R" in this case).
\end{thebibliography}
\end{document}